\documentclass[journal]{IEEEtran}

\usepackage{setspace}
\usepackage[noadjust]{cite}
\usepackage{multirow}
\usepackage{dsfont}
\usepackage[latin1]{inputenc}
\usepackage{times}
\usepackage [english]{babel}
\usepackage [T1]{fontenc}
\usepackage [latin1]{inputenc}
\usepackage[dvips]{graphicx}
\usepackage{grffile}
\usepackage{amssymb,amsmath,amsthm, amsfonts}
\usepackage{xspace}
\usepackage{cases}
\usepackage{lineno}
\usepackage{mathrsfs}

\renewcommand{\figurename}{Fig.}

\ifCLASSOPTIONcompsoc
\usepackage[tight,normalsize,sf,SF]{subfigure}
\else
\usepackage[tight,footnotesize]{subfigure}
\fi

\usepackage{algpseudocode}	
\usepackage{algorithm}
\usepackage[nolist]{acronym}

\renewcommand{\algorithmicrequire}{{\tt \textbf{Input:}}}

\usepackage[usenames,dvipsnames]{color}

\usepackage{pifont}

\newtheorem{lemma}{Lemma}
\newtheorem{corollary}{Corollary}
\newtheorem{definition}{Definition}

\theoremstyle{remark}
\newtheorem{remark}{Remark}

\usepackage{array}
\usepackage[normalem]{ulem} 
\usepackage{pgfplots}
\usepackage{tikz}
\usetikzlibrary{arrows}
\usetikzlibrary{shapes.multipart}
\usetikzlibrary{plotmarks}	
\usetikzlibrary{automata,positioning}
\usetikzlibrary{calc}

\pgfplotsset{
cycle list={%
{cyan,mark=diamond},
{brown,mark=square},
{red,mark=o},
{blue,mark=triangle},
{black,mark=otimes},
{magenta,mark=*},
},
every axis/.append style={%
enlarge x limits = false,
ylabel near ticks,
xlabel near ticks,
legend cell align=left,
font=\scriptsize,
semithick
},
}
\begin{document}

\newcommand{\fileDemDeadline}[1][k]{\ensuremath{t_{#1}}\xspace}

\def \ie{, i.e.,\xspace}
\def \eg{, e.g.,\xspace}
\def \cf{, c.f.,\xspace}
\def \wrt{\ac{w.r.t.}\xspace}
\def \etal{et al.\xspace}

\newcommand{\y}[1][i]{y_{#1}}
\newcommand{\pS}[1][k,n]{P_{#1}}
\newcommand{\x}[1][i]{x_{#1}}
\newcommand{\n}[1][i]{w_{#1}}
\newcommand{\s}[1][i]{s_{#1}}
\newcommand{\noiseVar}[1][i]{\sigma^2}

\def\mb{\mathbf}
\def\bs{\boldsymbol}
\def\bb{\mathbb}
\def\ds{\mathds}
\def\mc{\mathcal}
\def\ss#1{{\sf #1}}

\def\esc#1{\textrm{#1}}
\def\veci#1#2{\esc{#1}_{#2}}
\def\rveci#1#2{\uppercase{#1}_{#2}}
\def\rvecri#1#2{\lowercase{#1}_{#2}}

\def\vec#1{\mb{#1}}
\def\rvec#1{\bs{\uppercase{#1}}} 
\def\rvecr#1{\bs{\lowercase{#1}}} 
\def\vecs#1{\bs{#1}}

\def\mat#1{\mb{\uppercase{#1}}}
\def\mats#1{\bs{#1}}

\def\dim{n}
\def\dimp{m}
\def\dimpp{p}

\renewcommand{\Re}{\mathds{R}}
\def\R{\ds{R}}
\def\C{\ds{C}}
\def\N{\ds{N}}
\def\F{\R}
\def\RS{\R} 
\def\SymM#1{\ds{S}^{#1}}
\def\PSD#1{\SymM{#1}_+}
\def\PD#1{\SymM{#1}_{++}}

\def\Tr{\ss{Tr}}
\def\var{\ss{var}}
\def\det{\ss{det}}
\def\diag{\ss{diag}}
\def\Diag{\ss{Diag}}
\def\sign{\ss{sign}}
\def\log{\ss{log}}
\def\rank{\ss{rank}}
\def\pinv{\textrm{\footnotesize{+}}}
\def\H{\mathrm{H}}
\def\T{\ss{T}}
\def\d{\textnormal{\textrm{d}}}
\def\D{\ss{D}}
\def\v{\ss{vec}}

\newcommand{\lev}{\operatorname{lev}}
\newcommand{\eliminf}{\operatorname{e-lim\,inf}}
\newcommand{\elimsup}{\operatorname{e-lim\,sup}}
\newcommand{\elim}{\operatorname{e-lim}}
\newcommand{\argmin}{\operatorname{argmin}}
\newcommand{\inte}{\operatorname{int}}
\newcommand{\maximize}{\mathop{\operatorname{maximize}}}
\newcommand{\minimize}{\mathop{\operatorname{min}}}
\newcommand{\st}{\operatorname{s.t.}}

\begin{acronym}[OFDMA]
\acro{AP}{Access Point}
\acro{AWGN}{Additive White Gaussian Noise}
\acro{BCC}{Battery Capacity Constraint}
\acro{BER}{Bit Error Rate}
\acro{BWF}{Boxed Water-Flowing}
\acro{CSI}{Channel State Information}
\acro{CWF}{Classical Waterfilling}
\acro{DCC}{Data Causality Constraint}
\acro{DSL}{Digital Subscriber Line}
\acro{D2D}{Device-to-Device}
\acro{DWF}{Directional Water-Filling}
\acro{EH}{Energy Harvesting}
\acro{EBS}{Empty Buffers Strategy}
\acro{ECC}{Energy Causality Constraint}
\acro{FSA}{Forward Search Algorithm}
\acro{GNE}{Generalized Nash Equilibrium}
\acro{ICT}{Information and Communications Technology}
\acro{ICTs}{Information and Communications Technologies}
\acro{IFFT}{Inverse Fast Fourier Transform}
\acro{KKT}{Karush Kuhn Tucker}
\acro{i.i.d.}{independent and identically distributed}
\acro{ISCA}{Iterative Smooth and Convex approximation Algorithm}
\acro{ISO}{International Standards Organization}
\acro{ISS}{Incremental Slot Selection}
\acro{IDWF}{Iterative Directional Waterfilling}
\acro{IDWFA}{Iterative Directional Waterfilling Algorithm}
\acro{IV}{indicator variable}
\acro{IWFA}{Iterative Waterfilling Algorithm}
\acro{LCA}{Local Caching Algorithm}
\acro{LRU}{Least Recently Used}
\acro{LT}{Luby Transform}
\acro{MAC}{Medium Access Control}
\acro{MAP}{Maximum a Posteriori}
\acro{MBS}{Macro Base Station}
\acro{MFSK}{Multiple Frequency-Shift Keying}
\acro{MGSS}{Maximum Gain Slot Selection}
\acro{MI}{Mutual Information}
\acro{MIMO}{Multiple-Input Multiple-Output}
\acro{ML}{Maximum Likelihood}
\acro{MM}{Majorize and Minimize}
\acro{MMSE}{Minimum Mean-Square Error}
\acro{MUI}{MultiUser Interference}
\acro{NCP}{Nonlinear Complementary Problem}
\acro{NDA}{Non Decreasing water level Algorithm}
\acro{NE}{Nash Equilibrium}
\acro{OFDM}{Orthogonal Frequency Division Multiplexing}
\acro{OFDMA}{Orthogonal Frequency Division Multiple Access}
\acro{OSI}{Open System Interconnection}
\acro{PbP}{Pool-by-Pool}
\acro{PDCA}{Pre-Downloading Caching Algorithm}
\acro{QoS}{Quality of Service}
\acro{RF}{Radio Frequency}
\acro{SCA}{Successive Convex Approximation}
\acro{SINR}{Signal to Interference plus Noise Ratio}
\acro{SSA}{Successive Smooth Approximation}
\acro{SBS}{Small Base Station}
\acro{SISO}{Single-Input Single-Output}
\acro{SNR}{Signal to Noise Ratio}
\acro{SVD}{Singular Value Decomposition}
\acro{TDMA}{Time Division Multiple Access}
\acro{UPA}{Uniform Power Allocation}
\acro{VI}{Variational Inequality}
\acro{V-BLAST}{Vertical-Bell Laboratories Layered Space-Time}
\acroplural{VI}[VIs]{Variational Inequalities}
\acro{WEHN}{Wireless Energy Harvesting Node}
\acro{WER}{Word Error Rate}
\acro{WF}{WaterFilling}
\acro{WFlow}{Water-Flowing}
\acro{w.r.t.}{with respect to}
\acro{WSNs}{Wireless Sensor Networks}
\acro{WSS}{Weighted Slot Selection}
\end{acronym}

\graphicspath{{Figures/}}

\IEEEoverridecommandlockouts


\title{Wireless Content Caching for Small Cell and D2D Networks}
\author{ 
Maria Gregori,
Jesús Gómez-Vilardeb\'o, Javier Matamoros and Deniz Gündüz 
\thanks{Maria Gregori, Jesús Gómez-Vilardeb\'o, and  Javier Matamoros are with the Centre Tecnològic de Telecomunicacions
de Catalunya (CTTC), 08860 Barcelona, Spain (e-mails: \{maria.gregori, jesus.gomez, javier.matamoros\}@cttc.cat).
Deniz Gündüz is with the Imperial College of London, UK (e-mail: d.gunduz@imperial.ac.uk).}
\thanks{This work is partially supported by the EC-funded project NEWCOM\# (n.318306), by the Spanish Government through the projects INTENSYV (TEC2013-44591-P) and E-CROPs (PCIN-2013-027) in the framework of the ERA-NET CHIST-ERA program, and by the Catalan Government (2014 SGR 1567).} \vspace{-1cm}}

\maketitle

\begin{abstract}
The fifth generation wireless networks must provide fast and reliable connectivity while coping with the ongoing traffic growth. It is of paramount importance that the required resources, such as energy and bandwidth, do not scale with traffic. 
While the aggregate network traffic is growing at an unprecedented rate, users tend to request the same popular contents at different time instants. Therefore, caching the most popular contents at the network edge is a promising solution to reduce the traffic and the energy consumption over the backhaul links. In this paper, two scenarios are considered, where caching is performed either at a small base station, or directly at the user terminals, which communicate using \ac{D2D} communications. In both scenarios, joint design of the transmission and caching policies is studied when the user demands are known in advance. This joint design offers two different caching gains, namely, the \textit{pre-downloading} and \textit{local caching gains}. It is shown that the finite cache capacity limits the attainable gains, and creates an inherent tradeoff between the two types of gains. In this context, a continuous time optimization problem is formulated to determine the optimal transmission and caching policies that minimize a generic cost function, such as energy, bandwidth, or throughput. The jointly optimal solution is obtained by demonstrating that caching files at a constant rate is optimal, which allows to reformulate the problem as a finite-dimensional convex program.
The numerical results show that the proposed joint transmission and caching policy dramatically reduces the total cost, which is particularised to the total energy consumption at the \ac{MBS}, as well as to the total economical cost for the service provider, when users demand economical incentives for delivering content to other users over the D2D links.
\end{abstract}


\newcommand{\Ts}{\ensuremath{T_s}\xspace}
\newcommand{\fileAlph}[1][j]{\ensuremath{\mathds{F}}\xspace}
\newcommand{\file}[1][j]{\ensuremath{f_{#1}}\xspace}
\newcommand{\sizeFile}[1][j]{\ensuremath{l_{#1}}\xspace}

\newcommand{\indicatorUser}[1][j,n]{\ensuremath{\delta_{u}(#1)}\xspace}
\newcommand{\indicatorAll}[1][j,n]{\ensuremath{\sigma_{u}(#1)}\xspace}
\newcommand{\rhoFunction}[1][j,n]{\ensuremath{\vecs \rho}\xspace}

\newcommand{\serving}[1][t]{\ensuremath{\vec{s}(#1)}\xspace}
\newcommand{\servingUser}[1][u]{\ensuremath{s_{#1}}\xspace}
\newcommand{\servingNU}[1][nu]{\ensuremath{s_{#1}}\xspace}

\newcommand{\caching}[1][t]{\ensuremath{\vec c(#1)}\xspace}
\newcommand{\cachingP}[0]{\ensuremath{\vec c}\xspace}
\newcommand{\cachingUser}[1][t,u]{\ensuremath{c(#1)}\xspace}
\newcommand{\cached}[1][nu]{\ensuremath{q_{#1}}\xspace}
\newcommand{\cachedAll}[1][nu]{\ensuremath{\vec q}\xspace}

\newcommand{\demandUser}[1][t]{\ensuremath{d_{u}(#1)}\xspace}
\newcommand{\demandNU}[1][nu]{\ensuremath{d_{#1}}\xspace}

\newcommand{\rate}[1][t]{\ensuremath{\vec r}\xspace}
\newcommand{\rateMBS}[1][(t)]{\ensuremath{\vec r_{M}#1}\xspace}

\newcommand{\rateUser}[1][u]{\ensuremath{r_{#1}}\xspace}
\newcommand{\rateEpoch}[1][n]{\ensuremath{r_{#1}}\xspace}
\newcommand{\departure}[1][t]{\ensuremath{D(#1, r)}\xspace}

\newcommand{\optimalDeparture}[1][t, r^{\star}]{\ensuremath{D^\star(#1)}\xspace}
\newcommand{\optimalDepartureDagger}[1][t]{\ensuremath{D^\dagger(#1)}\xspace}
\newcommand{\optimalCaching}[1][(t)]{\ensuremath{\vec{c}^\star #1}\xspace}
\newcommand{\optimalCachingUser}[1][t,u]{\ensuremath{c^\star ({#1}) }\xspace}

\newcommand{\ratesDtoD}[1][u]{\ensuremath{ \vec{\hat r}\xspace}}
\newcommand{\ratesDtoDUser}[1][u]{\ensuremath{ \vec{\hat r_{#1}}}\xspace}
\newcommand{\rateUserUser}[2][t, u']{\ensuremath{\hat r_{#2} (#1)}\xspace}
\newcommand{\rateUserUserOptimal}[2][u]{\ensuremath{b^{\star}_{#1}(#2)}\xspace}
\newcommand{\rateUserUserDiscrete}[2][u]{\ensuremath{b_{#1}(#2)}\xspace}
\newcommand{\cachingDtoDUU}[2][t, u']{\ensuremath{c_{#2} (#1)}\xspace}
\newcommand{\cachingDtoDUUOptimal}[2][t, u']{\ensuremath{c^{\star}_{#2} (#1)}\xspace}
\newcommand{\departureUser}[1][t]{\ensuremath{D_{u}(#1, \rateUser)}\xspace}
\newcommand{\departureUserOptimal}[1][t]{\ensuremath{D^{\star}_{u}(#1, \rateUser)}\xspace}
\newcommand{\cachingDtoDuser}[1][u]{\ensuremath{\vec{c}_{#1}}\xspace}
\newcommand{\cachedDtoD}[2][u]{\ensuremath{q_{#1}(#2)}\xspace}
\newcommand{\cachedOptimal}[2][u]{\ensuremath{q^{\star}_{#1}(#2)}\xspace}

\renewcommand{\figurename}{Fig.}

\usetikzlibrary{positioning,calc}

\acresetall

\section{Introduction}
\label{sec_Intro}

Wireless traffic has  experienced a tremendous growth in the last years due to the wide spread use of  hand-held devices connected to the Internet\eg mobile phones, tablets, etc.
This traffic increase is expected to continue steadily in the coming years; for example,
more than 127 exabytes of worldwide mobile  traffic is forecasted for the year 2020 \cite{UMTStraffic_2011}. 
Video traffic is the major data source due to the growing success of on-demand video streaming services \cite{UMTStraffic_2011}.
Traffic resulting from video on-demand services exhibits the \textit{asynchronous content reuse} property \cite{Mingyue_D2D_2013}, according to which a few popular files, requested by users at different times (as opposed to television broadcasting services), account for most of the data traffic.

To cope with this growing traffic requirements, lots of efforts have been devoted towards the definition of the  fifth generation of cellular communication systems (5G), which is expected to be operative by 2020. 
The 5G system  must provide fast, flexible, reliable, and sustainable wireless connectivity, while supporting the growing mobile traffic. 
\ac{D2D} communications, small cell densification, millimeter wave, and massive MIMO are currently investigated as main enabling technologies for its success.

Small cell densification refers to the deployment of a large number of \acp{SBS} with different cell sizes (micro, pico, and femtocells) allowing a larger spatial reuse of the resources.
The major drawback of cell densification is that the traffic that can be served by an \ac{SBS} is limited by the capacity of the backhaul link, which provides connection to the core network.
This link is preferably wireless for various reasons such as, rapid deployment, self-configuration, and cost.
However, wireless backhaul connections  entail limited  capacity and significant energy consumption (due to its relatively long range).

Caching the most popular contents at the network edge has been  proposed 
in \cite{Goebbels_PIMRC_2008} to increase connectivity,  in \cite{Dehghan_infocom_2015} to reduce the delay, and in  \cite{golrezaei_femtocaching_2013}
to alleviate the backhaul link congestion  and to reduce its energy consumption.
Video traffic  (e.g., popular Youtube videos) is especially suitable to be cached since it requires high data rates and exhibits the aforementioned asynchronous content reuse property.
The contents can be cached either at \acp{SBS} equipped with a cache memory (also coined as ``femtocaching'') \cite{golrezaei_femtocaching_2013, Bastug_ProactiveCaching_2013, poularakis_multicast_2014,  poularakis_Algorithms_2014,   Blasco_ISIT_2014, pantisano_cache_2014, MaddahAli_FundamentalLimits_2014}, or directly at the users' devices  \cite{Mingyue_D2D_2013, Golrezaei_BSD2D_2013, Ji_order_2015}. 
The users can exchange the cached content through \ac{D2D} communications \cite{Ji_fundamental_2014}, which
allows  direct communication between nearby mobile users.
In practice, due to limited cache and energy resources, 
users are unwilling to serve data over the D2D links unless they obtain incentives (e.g., economical) from the operator \cite{PengLi_incentiveD2D}.

In a popular approach to wireless caching, \cite{MaddahAli_FundamentalLimits_2014, Ostovari_WCNC_2013, poularakis_multicast_2014}, the system design is performed in two separated phases. 
First, in the \textit{content placement} phase, each  cache is filled with  appropriate data, exploiting periods of time in which the network is not congested.
Then, in the \textit{delivery phase}, the non-cached contents are transmitted when requested by users.
In this setup, two types of caching gains  have been identified, namely, the \textit{local} and \textit{global caching gains} \cite{MaddahAli_FundamentalLimits_2014}.
On the one hand, the \textit{local caching gain} is obtained when a requested file is locally available in the  cache (either at the SBS or  at the users) by serving this file  from the cache without connecting to the \ac{MBS}.
This reduces the traffic in the wireless backhaul link \cite{poularakis_multicast_2014} and improves the quality of experience \cite{MaddahAli_FundamentalLimits_2014}. 
On the other hand, the \textit{global caching gain} is obtained by multicasting network-coded information in the delivery phase \cite{MaddahAli_FundamentalLimits_2014, Ostovari_WCNC_2013}.
However, this underlying separation between the caching (content placement) and transmission (delivery) phases has two limiting assumptions: 
i) the content placement phase is cost-free (e.g., in terms of energy or bandwidth); 
and ii) cache content is never updated during the delivery phase.
As a result, the benefits of proactive caching are inherently limited.

In this work, we consider a different approach to wireless caching. 
In particular, we consider that the cache is initially empty, and it is dynamically filled with contents\ie we combine the content placement and delivery phases.
This approach still allows  to pre-download data over low-traffic periods;
however, we now account for the cost of downloading these contents.
As a result, an additional caching gain is obtained, which we call  \textit{pre-downloading gain}.
Essentially, the \textit{pre-downloading caching gain} is obtained when the cache is used to pre-download data, which can be beneficial  
to
avoid non-favourable channel conditions, and 
to equalize the rate in the backhaul link, improving its energy efficiency, and reducing its peak load.
In this context, 
the authors of \cite{Sadr_Anticipatory_2013} and \cite{Gungor_proactiveCaching_2014} derive
caching and 
 transmission policies that minimize the  bandwidth  and energy consumption, respectively.
These works assume that the cache is solely used to pre-download content for a single user;
thus, content is removed from the cache as soon as it is consumed by the user, ignoring any possible future requests. 
Consequently, the policies in \cite{Sadr_Anticipatory_2013} and \cite{Gungor_proactiveCaching_2014} only exploit the pre-downloading caching gain.
To the best of our knowledge, this is the first work that proposes \textit{jointly} optimal transmission 
 and  caching strategies by accounting for both the local and  pre-downloading caching gains.

To fully exploit the aforementioned gains, efficient cache management policies must be designed by taking into account, among others, the stochastic but predictable nature of users' demands.
The cache management policies can be classified into two groups according to the prior knowledge of the different system parameters (users' requests, channel state information, etc.):
(i) \textit{offline} caching policies that assume non-causal and complete knowledge of these parameters, e.g., \cite{Sadr_Anticipatory_2013, Draxler_Anticipatory_2013, Gungor_proactiveCaching_2014};
and (ii) \textit{online} caching policies that consider only causal or probabilistic knowledge of these parameters, e.g., \cite{Blasco_ISIT_2014, Pedarsani_2015, Ostovari_WCNC_2013, Bastug_ProactiveCaching_2013,poularakis_multicast_2014,poularakis_Algorithms_2014,  pantisano_cache_2014}.
The characterization of the optimal offline policy is extremely useful because:
i) it  serves as a theoretical  bound on the performance achievable by any online policy; and ii)
it can be instrumental in designing low-complexity near-optimal online policies.
Finding the optimal online policy is extremely challenging since the cache management problem is usually a hard combinatorial problem.
As a result several works have resorted to heuristic algorithms \cite{Bastug_ProactiveCaching_2013, poularakis_multicast_2014}.

In contrast to previous literature, the aim of this paper is to study the \textit{jointly optimal} transmission and caching policies by taking into account both the local and pre-downloading caching gains under two different scenarios.
The first scenario considers a caching \ac{SBS} that serves the demands from multiple users. 
When the users' demand is not locally available at the SBS, the SBS downloads the content from an MBS through a wireless backhaul link. 
We 
addressed this scenario in  \cite{Gregori_ISIT_15} assuming 
 that the SBS serves the users in a time division fashion;
in this paper, we allow the SBS to serve multiple users simultaneously.
In the second scenario, we consider that the MBS directly serves demands from users, which can cache the received data proactively, and
later cooperate with other users through  D2D communications.
The key difference between the two scenarios is that, in the former, the cache is centralized at the SBS, whereas in the later, it is distributed across users.
The main contributions of the paper are summarized next:
\begin{itemize}
\item For the two scenarios mentioned above, we study the joint design of the optimal transmission and
caching policies by formulating a continuous time optimization problem aimed at  minimizing a generic cost function (e.g., the energy, throughput, or bandwidth requirement).
\item For the first scenario, where caching is performed at the SBS: 
(i) we show that, within each time slot, it is optimal to cache data at a constant rate, which permits reformulating the problem as a convex program;
(ii) we solve this convex problem by means of dual decomposition and propose a subgradient algorithm to obtain the optimal dual variables;
and (iii) we derive the structure of the optimal transmission power at the MBS and  the caching policy at the SBS.
\item For the second scenario, where information is cached at the users and shared through D2D communications:
(i) we show that, within each time slot, each user should cache data at a constant rate;
(ii)  we show that each user should transmit the files in the D2D links  at a constant rate;
and (iii) we reformulate the problem as a convex optimization problem.
\item Finally, the two scenarios are compared through numerical simulations.
 Fist, we 
compare the performance of a centralized cache with a distributed one and assess the impacts of pre-downloading and local caching gains in each scenario. 
Second, we evaluate how the cost of the MBS increases with the economical incentives requested by users for transmitting data over the D2D links.
\end{itemize}

The remainder of the paper is structured as follows. 
Section \ref{sec:SBS} focuses on the first scenario, where caching is performed at an SBS. 
In particular, the system model is presented in Section \ref{sec:sysModelFormulation};
the optimal transmission strategy is derived for a fixed caching policy in Section \ref{sec:SBSrepresentation}; 
and the problem is solved in Section \ref{sec_FormulationLinear}.
Section \ref{sec:D2D} is devoted to the second scenario where caching is performed at the user terminals.
The system model for this scenario is introduced in Section \ref{sec:D2D_model} and the resulting problem is solved in Section \ref{sec_FormulationLinearD2D}.
Section \ref{sec_Results} presents the numerical results.
Finally, the paper is concluded in Section \ref{sec_Conclusions}.


\textbf{Notation:}
Vectors and vector valued functions are denoted by lower case boldface letters\ie $\vec v$ and $\vecs \rho(\vec v)$, respectively. 
$(\vec v_u)_{u=1}^{U}$ defines a  column vector obtained by stacking the column vectors $\vec v_1, \dots, \vec v_U$ and
$[\vec v]_{k}$ returns the $k$-th element of the vector $\vec v$.
Symbol $\preceq$ denotes the component-wise ``smaller than or equal to'' inequality.
Finally, $[x]^+ \triangleq \max\{0, x\}$.

\section{\ac{SBS} caching for 5G networks}
\label{sec:SBS}

\vspace{-0.5em}
\subsection{System model and problem formulation}
\label{sec:sysModelFormulation}

\begin{figure}
\centering
\includegraphics[width=\columnwidth]{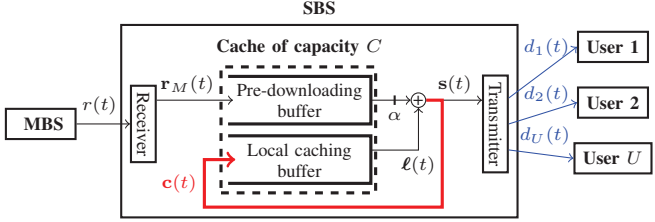}
\caption{System model when caching is performed at an SBS that serves $U$ users.}
\label{fig:systemModelSBS}
\end{figure}

As depicted in Fig. \ref{fig:systemModelSBS}, we consider $U$ users served by an \ac{SBS}. 
The SBS has a finite cache memory of capacity $C$ units, and is connected through a  wireless backhaul channel to an \ac{MBS}, which has access to the core network.
We assume that the MBS and the SBS operate in different frequency bands; thus, no interference is produced between the two.
We define 
$T$ as the optimization time horizon consisting of $N$ time slots of duration $\Ts$ each.
In each time slot, each user, $u$, $u = 1, \dots, U$, requests one file from the set
of all possible files  $\fileAlph=\{\file[0], \dots , \file[F]\}$.
We define $\sizeFile$ as the length (in data units) of  file \file,  $j = 0, \dots, F$.
File \file[0] has length  $\sizeFile[0] = 0$ and represents
slots 
 without requests.
Similarly to \cite{Draxler_Anticipatory_2013}, 
we fix the duration of 
each file in the set \fileAlph to the duration of one time slot, $\Ts$.\footnote{Note that any generic file can be partitioned into smaller files to meet the requirement of having the same duration $\Ts$.}
File \file is consumed at 
 a constant rate $\sizeFile / \Ts$  by the users.

As shown in Fig. \ref{fig:systemModelSBS}, 
data is transmitted by the MBS at a rate $r(t)$. 
The \ac{SBS} receives this information and separates the streams associated to  different user requests, obtaining the rate vector
$\rateMBS = (\rateUser(t))_{u=1}^{U}$ with
$\rateUser(t)$ being the rate associated to user $u$. 
The downloaded data, $r(t)= \sum_{u=1}^U r_u(t)$,
 is then stored at the \ac{SBS} cache until it is served to the users (which, without loss of generality, can happen immediately).
The SBS has a demand rate denoted by \serving to satisfy the users'  demand rates, $\demandUser$, $\forall u$.
As it will be explained later, the  SBS demand rate is obtained as the sum of the users'  demand rates, $\demandUser$, after removing multiple demands for the same file within the same slot.
When serving a content to a user, the SBS either deletes or locally caches it. 
This is dictated by the local caching rate $\caching$.
Notice that the cache is represented with two different virtual buffers, namely,
the pre-downloading and local caching buffers.
This representation with virtual buffers
allows us to distinguish 
between the cached data that is downloaded in advance from the MBS  from  the locally cached data that is used to reduce future requests from the MBS.
In this context, 
we define the vector $\vecs{\ell}(t)\triangleq (\ell_{u}(t))_{u=1}^{U}$
 whose $u$-th component $\ell_{u}(t)$ denotes 
  the rate at which  data is removed from the local caching buffer to reduce the demand at time $t$ from the MBS
associated to $u$-th user request\footnote{
In  Fig. \ref{fig:systemModelSBS} we represent the locally cached data as a feedback link from  the output to the input.
Note that the data removed from the local caching buffer, $\vecs{\ell}(t)$, can be instantaneously cached again if dictated by the local caching rate \caching (implying that in practice the content is not removed from the cache).
}.
In the sequel, we provide formal definitions for $\demandUser$, \serving, and \caching.
As in \cite{Sadr_Anticipatory_2013, Gungor_proactiveCaching_2014}, we assume a known demand profile (i.e, offline approach, see Section \ref{sec_Intro});
accordingly, we assume that the demand variables  $\demandUser$ and \serving are known for the period $[0,T]$.

We define $\indicatorUser$ as the \textit{user request indicator variable} that takes value $1$ when  user $u$ requests  file $\file$ in slot $n$,  $n = 1, \dots, N$, and  $0$, otherwise.
Since each user requests one file from \fileAlph per slot, we have $\sum_{j=0}^{F}\indicatorUser = 1$, $\forall u,n$.

\begin{definition}[User demand rate]
The demand rate of user $u$, $\demandUser \geq 0$, $t\in[0, T]$, is the rate at which 
user $u$ requests data from the  \ac{SBS}\ie
$\demandUser \triangleq \sum_{n=1}^{N} \demandNU  \operatorname{rect} ( (t - (n - 1/2)\Ts) /\Ts   )$,
where \demandNU denotes the demand rate of the $u$-th user at the $n$-th time slot\ie
$\demandNU = \sum_{j=0}^{F}  \indicatorUser \sizeFile / \Ts$; and
$\operatorname{rect}((t- a)/b)$ stands for the rectangular function centered at $a$ with duration $b$. 
\end{definition}

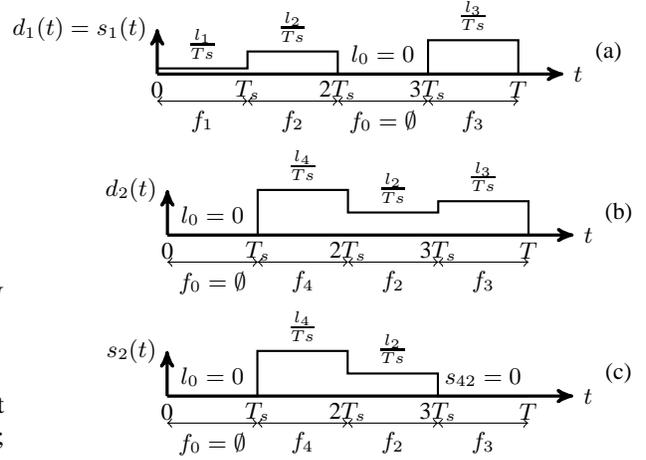
\begin{figure}
\centering
\hspace{-1.5cm}
\begin{tikzpicture}[
    xscale=0.3,
    yscale=0.3,
    axis/.style={very thick, ->, >=stealth'},
    important line/.style={thick},
    dashed line/.style={dashed, thin},
    pile/.style={thick, ->, >=stealth', shorten <=2pt, shorten
    >=2pt},
    every node/.style={color=black,font=\small}
    ]
    \draw[axis] (0,0)  -- (18,0) node(xline)[right]
        {$t$};
    \draw[axis] (0,0)  -- (0,2) node(yline)[left]
        {$d_1(t)= s_1(t)$};

   \draw[important line] (0,0.25) -- (4,0.25) node[midway, above] {$\frac{\sizeFile[1]}{Ts}$}  --(4,1)-- (8,1) node[midway, above] {$ \frac{\sizeFile[2]}{Ts}$} --(8,0) -- (12,0) node[midway, above] {$\sizeFile[0]=0$}  --(12,1.5) -- (16,1.5) node[midway, above] {$\frac{\sizeFile[3]}{Ts}$} --(16,0);

    \node at (0,-0.7){0};
    \node at (4,-0.7){\Ts};
    \node at (8,-0.7){2\Ts};
    \node at (12,-0.7){3\Ts};
        	\node at (16,-0.7) {$T$};  
        
    \draw [<->] (0,-1.2)  --  (4,-1.2) node[midway, below] {\file[1]};;          
    \draw [<->] (4,-1.2)  --  (8,-1.2) node[midway, below]{\file[2]};
    \draw [<->] (8,-1.2)  --  (12,-1.2) node[midway, below] {$ \file[0] = \emptyset$};    
    \draw [<->] (12,-1.2)  --  (16,-1.2) node[midway, below] {\file[3]};
	
    \node at (20, 1) {(a)};
    
\end{tikzpicture}
\begin{tikzpicture}[
    xscale=0.3,
    yscale=0.3,
    axis/.style={very thick, ->, >=stealth'},
    important line/.style={thick},
    dashed line/.style={dashed, thin},
    pile/.style={thick, ->, >=stealth', shorten <=2pt, shorten
    >=2pt},
    every node/.style={color=black,font=\small}
    ]
    \draw[axis] (0,0)  -- (18,0) node(xline)[right]
        {$t$};
    \draw[axis] (0,0)  -- (0,2) node(yline)[left]
        {$d_2(t)$};

   \draw[important line] (0,0) -- (4,0) node[midway, above] {$\sizeFile[0]= 0$}  --(4,2)-- (8,2) node[midway, above] {$\frac{\sizeFile[4]}{Ts}$} --(8,1)-- (12,1) node[midway, above] {$ \frac{\sizeFile[2]}{Ts}$}  --(12,1.5) -- (16,1.5) node[midway, above] {$\frac{\sizeFile[3]}{Ts}$} --(16,0);

    \node at (0,-0.7){0};
    \node at (4,-0.7){\Ts};
    \node at (8,-0.7){2\Ts};
    \node at (12,-0.7){3\Ts};
	\node at (16,-0.7) {$T$};  
        
    \draw [<->] (0,-1.2)  --  (4,-1.2) node[midway, below] {$\file[0]= \emptyset$};;          
    \draw [<->] (4,-1.2)  --  (8,-1.2) node[midway, below]{\file[4]};
    \draw [<->] (8,-1.2)  --  (12,-1.2) node[midway, below] {$\file[2]$};    
    \draw [<->] (12,-1.2)  --  (16,-1.2) node[midway, below] {\file[3]};
	
    \node at (20, 1) {(b)};
    
\end{tikzpicture}
\begin{tikzpicture}[
    xscale=0.3,
    yscale=0.3,
    axis/.style={very thick, ->, >=stealth'},
    important line/.style={thick},
    dashed line/.style={dashed, thin},
    pile/.style={thick, ->, >=stealth', shorten <=2pt, shorten
    >=2pt},
    every node/.style={color=black,font=\small}
    ]
    \draw[axis] (0,0)  -- (18,0) node(xline)[right]
        {$t$};
    \draw[axis] (0,0)  -- (0,2) node(yline)[left]
        {$\servingUser[2](t)$};

 \draw[important line] (0,0) -- (4,0) node[midway, above] {$\sizeFile[0]= 0$}  --(4,2)-- (8,2) node[midway, above] {$\frac{\sizeFile[4]}{Ts}$} --(8,1)-- (12,1) node[midway, above] {$ \frac{\sizeFile[2]}{Ts}$}   --(12,0) -- (16,0) node[midway, above] {$\servingNU[42] =0$};


    \node at (0,-0.7){0};
    \node at (4,-0.7){\Ts};
    \node at (8,-0.7){2\Ts};
    \node at (12,-0.7){3\Ts};
	\node at (16,-0.7) {$T$};  
        
    \draw [<->] (0,-1.2)  --  (4,-1.2) node[midway, below] {$\file[0]= \emptyset$};;          
    \draw [<->] (4,-1.2)  --  (8,-1.2) node[midway, below]{\file[4]};
    \draw [<->] (8,-1.2)  --  (12,-1.2) node[midway, below] {$\file[2]$};    
    \draw [<->] (12,-1.2)  --  (16,-1.2) node[midway, below] {\file[3]};
	
    \node at (20, 1) {(c)};
    
\end{tikzpicture}

\vspace{-1em}
\caption{(a) and (b) denote the demand rates of users one and two, respectively. 
The values above the curves represent the value of \demandNU.
User 1 requests the files $\file[1]$, $\file[2]$, $\file[0]$, and $\file[3]$, and user 2 requests 
$\file[0]$, $\file[4]$, $\file[2]$, and $\file[3]$ in this order.
The user request indicator variable for user $1$ takes values $\delta_{1}(1,1) = \delta_{1}(2,2) =\delta_{1}(0,3) = \delta_{1}(3,4) =1$, and 
$0$, otherwise; similarly,
for user $2$,
$\delta_{2}(0,1) = \delta_{2}(4,2) =\delta_{2}(2,3) = \delta_{2}(3,4) =1$, and 
it is $0$, otherwise.
(c) shows the required  SBS demand rate for user $2$ (for user $1$, we have  $\servingUser[1](t) = d_1(t)$ as shown in (a)).
The values above the curves represent the value of \servingNU[n2].
Note that $\servingNU[n2] = \demandNU[n2]$ for all the slots except  the fourth one, where we have $\servingNU[42] =0$, because the two users request  file \file[3] in the fourth slot; thus, we have $\sigma_{2}(3,4) =0$.
} 
\label{fig:requests}
\end{figure}

Figs. \ref{fig:requests}(a)-(b) depict users' demand rates when two users are served from the SBS.
Note that if  a certain file is requested by multiple users at the same time slot (as in the fourth time slot in Fig. \ref{fig:requests}), these requests 
can be simultaneously handled by the \ac{SBS} without the need of downloading the same file multiple times from the \ac{MBS}.
Therefore, in order to determine the minimal
demand rate of the  \ac{SBS}, we must  account only once for simultaneous requests of the same file within one slot. 
Without loss of generality, we account for the request of the user with the smallest index $u$.
Accordingly,  we define the \textit{SBS request indicator variable}
$\indicatorAll$  that takes value $1$ for the user with the smallest index $u$ requesting file $\file$ in time slot $n$ (i.e., $\indicatorAll= 1$ if $\indicatorUser = 1$ and $u< u', \forall u'\neq u: \delta_{u'}(j,n) = 1$), and $0$, otherwise.

\begin{definition}[SBS demand rate]
The  demand rate of the \ac{SBS}
is
denoted by the vector  $\serving = (\servingUser(t))_{u=1}^{U}$, $t\in[0, T]$.
The $u$-th component of this vector, $\servingUser(t)\geq 0$, identifies the rate at which the data corresponding to the $u$-th user request must be
 available at the \ac{SBS} to fulfill this request.
 Thus, 
we have
$\servingUser(t) \triangleq \sum_{n=1}^{N} \servingNU \operatorname{rect} ( (t - (n - 1/2)\Ts) /\Ts   )$,
where \servingNU denotes the SBS demand rate at the $n$-th slot for the $u$-th user request, $\servingNU = \sum_{j=0}^{F}  \indicatorAll \sizeFile/ \Ts $. 
\end{definition}

Given the users' demands in Figs. \ref{fig:requests}(a)-(b), 
the associated SBS demand rates are shown in Figs. \ref{fig:requests}(a) and \ref{fig:requests}(c), respectively.
Note that if the SBS demand, \serving, is satisfied, then the SBS can serve all the user requests, $\demandUser, \forall u$.
In the remainder of this section, unless it is stated otherwise,
by a  user request
 we refer to the request seen by the SBS, $\servingUser(t)$, instead of the request on the user side, $\demandUser$.


\begin{definition}[Local caching rate]
The local caching rate is represented by the vector $\caching \triangleq (\cachingUser)_{u=1}^{U}$ whose $u$-th component, $\cachingUser$,
denotes the rate at which the \ac{SBS} caches the content associated to user $u$  at time $t$.
Thus, we have  
$0 \leq \cachingUser \leq \servingUser(t)$, $t\in[0, T]$.
\end{definition}


\begin{remark}
The variables in the system model must be able to indicate which portions of each file are cached at  time $t$.
Given a certain user request $u$ and  time instant $t$, we can identify the file being requested (through the SBS indicator variables).
Then, the cached portions of a file can be identified with the tuple  $\{\rateMBS, \caching, \serving\}$.
Note that we could have defined the 
vectors $\rateMBS, \caching,$  and $\serving$
in terms of the files instead of user requests (i.e., with dimensions  $F \times 1$ instead of $U \times 1$), which would 
simplify the identification of the cached files; however, this would dramatically increase the computational complexity of the  algorithms  proposed in the remainder of the paper as, in general, the number of available files, $F$, is several orders of magnitude larger than the number of users connected to the SBS, $U$. 
\end{remark}

Our aim is  to \textit{jointly} design the transmission policy at the \ac{MBS}, $r\triangleq  \{r(t)\}_{t=0}^{T}$, and the local caching policy at the \ac{SBS}, $\cachingP \triangleq  \{\caching\}_{t=0}^{T}$,
to minimize a generic cost function in the backhaul link, $\int_{0}^{T} g(r(\tau)) d\tau$, where  $g(r(t))$ denotes the instantaneous cost, which depends on the instantaneous transmission rate at the MBS.
As in \cite{zafer_calculus_2009}, we assume that the instantaneous cost function $g(\cdot)$ is time invariant, convex, increasing, continuously differentiable, and
$g(0) = 0$.
In the following, we give four examples of cost functions that satisfy these conditions:

\noindent
1) \textbf{Energy consumption minimization:}
If the objective is to minimize the total network energy consumption, then
the instantaneous cost is given by the instantaneous total power consumption, $p(t)$.  
In the case of Gaussian signaling, we have  $p(t)= g(r(t))= (\exp(r(t)) - 1)/h + P_c +  p_{S}(t)$, where $h$ denotes the channel gain, $P_c$ stands for the static circuitry consumption at the MBS, and $p_{S}(t)$ is the  SBS power consumption, which is known as it can be computed from the power-rate function at the SBS and  the users' demands.

\noindent
2) \textbf{Energy cost minimization:}
The instantaneous  power consumption above  has to be multiplied by the energy cost, $\xi_{MBS}$,
paid by the network operator to the electricity utility. Thus, the instantaneous cost function is $g(r(t))= \xi_{MBS}\cdot ((\exp(r(t)) - 1)/h + P_c +  p_{S}(t) )$.

\noindent
3) \textbf{Bandwidth minimization:}
 In this case, the cost function is given by the bandwidth-rate function, $w(t) =  g(r(t))= f^{-1}(r(t))$, obtained as the inverse of the rate-bandwidth function, $r(t) = f( w(t) )$. Again, in the case of Gaussian signaling, we have  $r(t) = f( w(t) )= w(t) \log(1 + Ph/w(t))$, where $P$ denotes the constant transmission power and $h$ stands for the channel gain.
 
\noindent
4) \textbf{Traffic minimization:}
To minimize the data transmitted by the MBS, we  obtain $g(r(t))= r(t)$.


As argued in the introduction, the cache offers two different  gains to reduce the cost in the backhaul link, namely, \textit{pre-downloading} and  \textit{local} caching gains. 
As shown in  Fig. \ref{fig:systemModelSBS}, the cache has two inputs:
(i) the pre-downloaded data from the MBS, which is controlled by the transmission policy at the \ac{MBS}, $r$, and contributes to the pre-downloading caching gain;
and (ii) the locally cached data, which is controlled by the local caching policy  at the \ac{SBS}, $\cachingP$, and contributes to the local caching gain. 
The design of $r$ and \cachingP is constrained by the cache size and the required demand rate at the SBS.
In the following, we define these constraints in terms of the cumulative transmitted data \cite{zafer_calculus_2009}.


\begin{definition}[Data departure curve]
The data departure curve, $\departure$, is the amount of total data served by the \ac{MBS} by time $t\geq 0$, and can be obtained from the  transmission policy, $r$, as $\departure \triangleq  \int_{0}^{t}  r (\tau)\d \tau$.
\end{definition}

Due to the  finite cache capacity,
an upper bound on 
 $\departure$ must be imposed to avoid data overflows from the SBS cache.
 This upper bound is imposed by the maximum data departure curve that, as defined next,
 increases as data is removed from the SBS cache. The rate at which data is removed from the cache at time $t$ is obtained as $ \sum_{u=1}^U  \servingUser(t) - \cachingUser[t,u]$.
\begin{definition}[Maximum data departure curve]
\label{def:maxD}
The maximum data departure curve, $B(t,\cachingP)$, limits the maximum amount of total data that can be transmitted by the MBS by time $t\geq 0$ such that no data overflow at the cache memory is generated.
Thus, it is given by $B(t,\cachingP) \triangleq C +   \int_{0}^{t} \sum_{u=1}^U \servingUser(\tau) - \cachingUser[\tau,u] \d \tau$ and depends on the caching policy $\cachingP$. 
\end{definition}

\begin{figure}
\centering

\begin{tikzpicture}[
    xscale=0.34,
    yscale=0.35,
    axis/.style={very thick, ->, >=stealth'},
    important line/.style={thick},
    dashed line/.style={dashed, thin},
    pile/.style={thick, ->, >=stealth', shorten <=2pt, shorten
    >=2pt},
    every node/.style={color=black,font=\small}
    ]
    \draw[axis] (0,0)  -- (18,0) node(xline)[right]
        {$t$};
    \draw[axis] (0,-1.50)  -- (0,3) node(yline)[left]
        {$[\rhoFunction(t,1)]_{1}$};

   \draw[important line] (0,-1.5) node[left]{-1} -- (8,-1.5)--(8,-1.2)  (8,0) -- (12,1.5)-- (12,0) (12,-1.2) -- (12,-1.5) -- (16,-1.5) --(16,-1.2);

    \node [left] (0,0){0};
    \node at (4,-0.7){\Ts};
    \node at (8,-0.7){2\Ts};
    \node at (12,-0.7){3\Ts};
        	\node at (16,-0.7) {$T$};  
        
    \draw [<->] (0,2)  --  (4,2) node[midway, above] {\file[1]};          
    \draw [<->] (4,2)  --  (8,2) node[midway, above]{\file[2] };
    \draw [<->] (8,2)  --  (12,2) node[midway, above] {$ \file[0] = \emptyset$};    
    \draw [<->] (12,2)  --  (16,2) node[midway, above] {\file[3]};

    \node[blue] at (-2,-3) {$[\rhoFunction(t,1)]_{2}$};
    \draw [<->,blue] (0,-3.5)  --  (8,-3.5) node[midway, above,blue] {$ -1$};
    \draw [<->,blue] (8,-3.5)  --  (12,-3.5) node[midway, above,blue] {$2$};    
    \draw [<->,blue] (12,-3.5)  --  (16,-3.5) node[midway, above,blue] {$-1$};               

	 \draw [dashed] (0,1.5) node[left]{\Ts} --  (12,1.5);

	\node at (8, -5) {(a)};

\end{tikzpicture}
\begin{tikzpicture}[
    xscale=0.34,
    yscale=0.35,
    axis/.style={very thick, ->, >=stealth'},
    important line/.style={thick},
    dashed line/.style={dashed, thin},
    pile/.style={thick, ->, >=stealth', shorten <=2pt, shorten
    >=2pt},
    every node/.style={color=black,font=\small}
    ]
    \draw[axis] (0,0)  -- (18,0) node(xline)[right]
        {$t$};
    \draw[axis] (0,-1.50)  -- (0,4.5) node(yline)[left,yshift= -0.5,xshift=1]
        {$[\rhoFunction(t,2)]_{1}$};

   \draw[important line] (0,-1.5) node[left]{-1} -- (8,-1.5) -- (8,-1.2) (8,0)-- (8,1.5) -- (12,3)-- (12,0) (12,-1.2) -- (12,-1.5) -- (16,-1.5) --(16,-1.2);

    \node [left] (0,0){0};
    \node at (4,-0.7){\Ts};
    \node at (8,-0.7){2\Ts};
    \node at (12,-0.7){3\Ts};
        	\node at (16,-0.7) {$T$};  
        
    \draw [<->] (0,3.5)  --  (4,3.5) node[midway, above] {$ \file[0] = \emptyset$};          
    \draw [<->] (4,3.5)  --  (8,3.5) node[midway, above]{\file[4] };
    \draw [<->] (8,3.5)  --  (12,3.5) node[midway, above] {\file[2]};    
    \draw [<->] (12,3.5)  --  (16,3.5) node[midway, above] {\file[3]};

	\node[blue] at (-2,-3) {$[\rhoFunction(t,2)]_{2}$};
    \draw [<->,blue] (0,-3.5)  --  (8,-3.5) node[midway, above,blue] {$ -1$};
    \draw [<->,blue] (8,-3.5)  --  (12,-3.5) node[midway, above,blue] {$1$};    
    \draw [<->,blue] (12,-3.5)  --  (16,-3.5) node[midway, above,blue] {$-1$};

	 \draw [dashed] (0,1.5) node[left]{\Ts} --  (12,1.5);
 	 \draw [dashed] (0,3) node[left]{$2\Ts$} --  (12,3);

	\node at (8, -5) {(b)};
    
\end{tikzpicture}

\vspace{-1em}
\caption{Representation of the functions $\rhoFunction(t,u)$ that maps a certain file request in the SBS to its previous occurrence in time ($t'=[\rhoFunction(t,u)]_{1}$) and user index ($u'=[\rhoFunction(t,u)]_{2}$). (a) corresponds to the first user and (b) to the second one.}
\label{fig:previousRequest}
\end{figure}
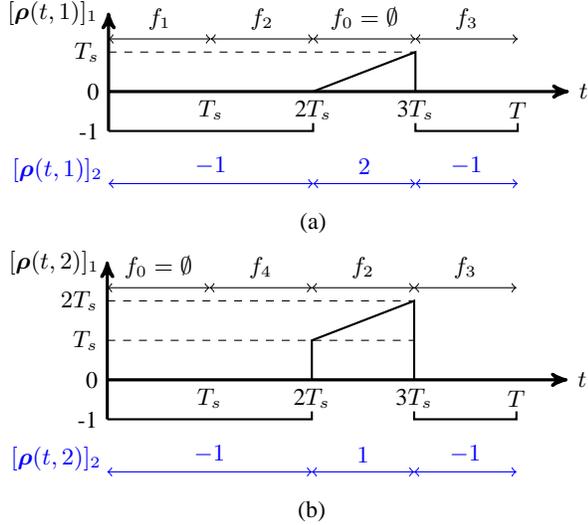

The lower bound on the data departure curve is given by the minimum amount of total data that must be  downloaded from the MBS to satisfy the SBS demand rate.
The \textit{net SBS demand rate} from the MBS (the demand rate at point $\alpha$ in Fig. \ref{fig:systemModelSBS}) is the requested data that is not available in the local caching buffer.
Consider that, at a certain time instant $t$,
user $u$ requests a file that had been previously requested by user $u'$ at time $t'$, $t'<t$.
Then,  
the net SBS demand rate at time $t$ of the $u$-th user request
is given by 
$\servingUser[u](t) - \ell_u(t)$, where, as mentioned earlier,  $\ell_u(t)$  denotes the rate at which data is removed from the local caching buffer to reduce the demand at time $t$ from the MBS.
Note that the rate $\ell_u(t)$  must be equal to the caching rate 
adopted during the previous request of the file requested by user $u$ at time $t$\ie $\ell_u(t)=  \cachingUser[t',u']$ (as otherwise data is unnecessarily downloaded from the MBS).
To compute the net SBS demand rate
for any user and time instant, we define the  vector function
$[t', u'] = \rhoFunction(t,u)$.
 This function returns the time instant $t'$ and the 
index  of the user, $u'$, which last requested the file requested by user $u$ at time $t$. 
If the file being requested  at time $t$ by user $u$
has not been requested previously, we set $\rhoFunction(t,u)=[-1,-1]$, and  define  $\cachingUser[-1,-1] \triangleq 0$ 
(since the files that have not been requested are not yet available at the \ac{SBS}).
The function $\rhoFunction(t,u)$ is depicted in Fig. \ref{fig:previousRequest} for the demand profile in Fig. \ref{fig:requests}. 
Using the function $\rhoFunction(t,u)$, the net SBS demand rate at time $t$ associated with  the $u$-th user request
is given by $\servingUser[u](t) - c(\rhoFunction(t, u))$.
Since non-causal knowledge of the user demands is available (offline approach), 
 the function $\rhoFunction(t,u)$ is known, $\forall t, u$.
Next we define the minimum data departure curve to satisfy the SBS demand.

\begin{definition}[Minimum data departure curve]
\label{def:minD}
The minimum data departure curve, $A(t,\cachingP)$, is the minimum amount of total data that must be transmitted by the \ac{MBS} by time $t\geq 0$ to satisfy the SBS demand, and depends on the caching policy, $\cachingP$\ie  $A(t,\cachingP) \triangleq
 \sum_{u=1}^U  \int_{0}^{t} \servingUser(\tau)- \ell_u (\tau) \d \tau =
 \sum_{u=1}^U  \int_{0}^{t} \servingUser(\tau)- \cachingUser[\rhoFunction(\tau, u)] \d \tau$.
\end{definition}

Bearing all the above in mind,  the problem is mathematically formulated as follows: 
\begin{subequations}
\label{eq:problem}
\begin{eqnarray}
\minimize_{ \{r(t), \caching \}_{t\in[0,T]}} &  \int_{0}^{T} g (r(\tau)) \d \tau &\\
\st 
  &\departure \leq B(t,\cachingP),& \quad  \forall t \in [0,T], \label{eq:c2} \\
    &\departure \geq A(t,\cachingP),& \quad \forall t \in [0,T],  \label{eq:c1}\\
  & r(t) \geq 0, \hfill & \quad \forall t \in [0,T], \label{eq:c3} \\
  &\vec 0 \preceq \caching \preceq \serving,& \quad \forall t \in [0,T], \label{eq:c4} 
\end{eqnarray}
\end{subequations}
where the constraint  \eqref{eq:c2} prevents  cache overflows, and \eqref{eq:c1} imposes the fulfillment of the users' demands. 
The constraints  \eqref{eq:c3} and  \eqref{eq:c4} guarantee feasible transmission and local caching rates.
Note that a feasible caching policy, $\cachingP$, must satisfy  $B(t,\cachingP)\geq A(t,\cachingP)$ for all $t \in [0,T]$, and any feasible data departure curve must lie within the tunnel between $B(t,\cachingP)$ and  $A(t,\cachingP)$. 
\begin{remark}
\label{Remark:cachingNextRequest}
In the problem formulation, we have assumed  that
 cached data can only be removed from the cache during the subsequent requests of the same data.
As a result, by caching data the net SBS demand rate will be  reduced.
We remark here that this assumption is without loss of optimality. 
Contrarily, consider a policy that caches a certain data content at time $t_1$, its subsequent request occurs at $t_2$, but the content is deleted at $t_3\in (t_1,t_2)$. As this content has to be downloaded again at $t_2$, this policy is unnecessarily using cache space in $(t_1,t_3)$.
\end{remark}

\begin{remark}
In realistic 5G scenarios,
 several SBSs will be served by the same MBS. This work considers that the MBS assigns orthogonal resources to each SBS 
and that the SBSs have non overlapping coverage areas. As a result, a problem of the form of \eqref{eq:problem} is obtained for each SBS. 
Further gains can be achieved by multicasting information to different SBSs, or by cooperation among SBSs with overlapping coverage areas \cite{golrezaei_femtocaching_2013, pantisano_cache_2014b}. However, this will inherently couple the design of the SBSs' caching policies, and is out of the scope of this work.
\end{remark}

\vspace{-1em}
\subsection{Optimal transmission strategy for a fixed caching policy}
\label{sec:SBSrepresentation}

\begin{figure}
\begin{tikzpicture}[
    xscale=0.4,
    yscale=0.22,
    axis/.style={very thick, ->, >=stealth'},
    important line/.style={thick},
    dashed line/.style={dashed, thin},
    pile/.style={thick, ->, >=stealth', shorten <=2pt, shorten
    >=2pt},
    every node/.style={color=black,font=\small}
    ]
    \draw[axis] (0,0)  -- (16,0) node(xline)[right] {$t$};
    \draw[axis] (0,0)  -- (0,7) node(xline)[left] {Data};

    \draw[red, important line] (0,0) -- (4,0.25)  -- (8, 3.25)  -- (12,4.25) --(16,5.75) ;
    
    \draw[blue, important line] (0,1) -- (4,1.25) node[midway, above, blue, sloped]{\servingNU[11]}  -- (8, 4.25) node[midway, blue, sloped, above, rotate=-10]{\servingNU[21]+ \servingNU[22]}-- (12,5.25)  node[midway, above,blue, sloped]{\servingNU[32]}  --(16,6.75)  node[midway, above,blue, sloped]{\servingNU[41]};

    \draw[dashed, important line] (0,0)-- (4,1.25)-- (8,3.25)--(16,5.75);

	\draw[<->] (8,0)--(8,3.25) node[right,midway]{\sizeFile[1]+\sizeFile[2]+\sizeFile[4]};
	\draw[dotted] (8,3.25)--(12,3.25);
	\draw[<->] (12,3.25)--(12,4.25) node[right,midway]{\sizeFile[2]};
	\draw[dotted] (12,4.25)--(16,4.25);
	\draw[<->] (16,4.25)--(16,5.75) node[right,midway]{\sizeFile[3]};

    \node at (0,-1){0};
    \node at (4,-1){\Ts};
    \node at (8,-1){2\Ts};
    \node at (12,-1){3\Ts};
	\node at (16,-1) {$T$};

    \draw [<->] (-0.5,0)  --  (-0.5,1);
	\node[align=center, left] at (-0.5,0.5){$C$};

	 \draw [blue, ->] (4,4) node[blue, above] {$B(t, \vec{\hat c})$}  --  (4,1.5);
 	 \draw [red, ->] (14, 2.5) node[red, below] {$A(t,  \vec{\hat c})$}  --  (14,5);
	 
	 	 	\node[align=center] at (8, -4){$(a)$};	 

\end{tikzpicture}
%
%
\begin{tikzpicture}[
    xscale=0.4,
    yscale=0.22,
    axis/.style={very thick, ->, >=stealth'},
    important line/.style={thick},
    dashed line/.style={dashed, thin},
    pile/.style={thick, ->, >=stealth', shorten <=2pt, shorten
    >=2pt},
    every node/.style={color=black,font=\small}
    ]
    \draw[axis] (0,0)  -- (16,0) node(xline)[right] {$t$};
    \draw[axis] (0,0)  -- (0,7) node(xline)[left] {Data};

    \draw[red, important line] (0,0) -- (4,0.25) node[midway, above, blue, rotate = 2, yshift=4]{\servingNU[11]} -- (8, 3.25) node[midway, above,blue, rotate = 20, yshift=4]{\servingNU[22]} -- (12,3.25) node[midway, above,blue, rotate = 10, yshift=4]{\servingNU[32]} --(16,4.75) node[midway, above,blue, rotate = 10, yshift=4]{\servingNU[41]};

    \draw[blue, important line] (0,1) -- (4,1.25) -- (8, 3.25)-- (12,4.25)--(16,5.75);

    \draw[dashed, important line] (0,0)-- (4,1.25)-- (8,3.25)--(16,4.75);

	\draw[<->] (8,0)--(8,3.25) node[right,midway]{\sizeFile[1]+\sizeFile[2]+\sizeFile[4]};
	\draw[dotted] (12,3.25)--(16,3.25);
	\draw[<->] (16,3.25)--(16,4.75) node[right,midway]{\sizeFile[3]};

    \node at (0,-1){0};
    \node at (4,-1){\Ts};
    \node at (8,-1){2\Ts};
    \node at (12,-1){3\Ts};
	\node at (16,-1) {$T$};

    \draw [<->] (-0.5,0)  --  (-0.5,1);
	\node[align=center, left] at (-0.5,0.5){$C$};

	 \draw [blue, ->] (4,4) node[blue, above] {$B(t, \vec{\tilde c} )$}  --  (4,1.5);
	 
	 \draw [blue, dotted, ->] (8,6.5) node[above] (cached) {\file[2] is cached}  --  (7.5,3.8);
 	 \draw [red,dotted, ->] (cached.south)   --  (11,3.2);
	 
 	 \draw [red, ->] (14, 2.5) node[red, below] {$A(t,  \vec{\tilde c})$}  --  (14,4);
	 
	 	 	\node[align=center] at (8, -4){$(b)$};
\end{tikzpicture}
\vspace{-1em}
\caption{Representation of the problem for two different caching policies. In this example, we have set $C = \sizeFile[2]$.}
\label{fig_probRepresentation}
\end{figure}
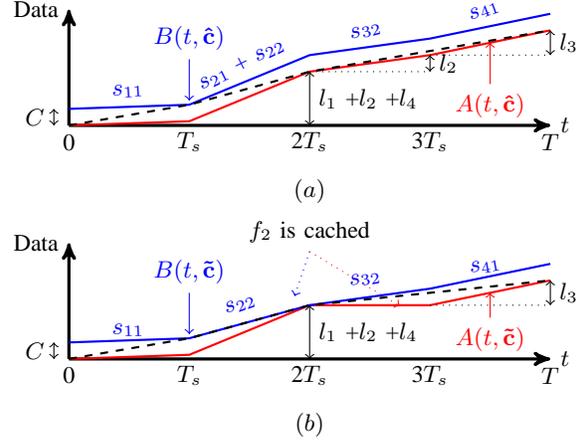

In this section, we  derive the optimal transmission strategy for a fixed caching policy.
Interestingly, when the local caching policy, $\cachingP$, is given, 
the problem in  \eqref{eq:problem} accepts an intuitive graphical representation.
For example, under the \ac{SBS} demand rate in Figs. \ref{fig:requests}(a) and \ref{fig:requests}(c), the problem is represented in Fig. \ref{fig_probRepresentation} for two different caching policies:

\textbf{Policy 1:}  The policy $\vec{\hat c}$ shown in Fig. \ref{fig_probRepresentation}(a) removes the data from the cache as soon as it is 
served to a user, ignoring any possible future requests for the same file\ie $\vec{\hat c} (t) = \vec 0$, $\forall t$.
 Consequently, it only exploits  the pre-downloading caching gain. This caching policy was proposed in \cite{Gungor_proactiveCaching_2014}. 
Observe that if  $\vec{\hat c} (t) = \vec 0$, $\forall t$, then 
there is a constant gap of $C$ between the lower and upper bounds\ie
$ B(t, \vec{\hat c} ) = C +  A(t, \vec{\hat c} )$ (c.f. Definitions \ref{def:maxD} and \ref{def:minD}). 
The optimal data departure curve exploits this gap by pre-downloading data.

\textbf{Policy 2:} The policy $\vec{\tilde c}$ shown in Fig. \ref{fig_probRepresentation}(b) caches the file $\file[2]$, when requested by user $1$ in the second time slot, thus anticipating the next request in the third slot by user $2$\ie $\tilde{c}(t,1)= \servingNU[21]$ if $t\in [\Ts, 2\Ts]$ and  $\tilde{c}(t,u)= 0$, otherwise. As a result, no data needs to be transmitted by the \ac{MBS} in the third slot.

\begin{lemma}[Constant rate transmission is optimal \cite{zafer_calculus_2009}]
\label{remark:constantRate}
Given a feasible caching policy $\cachingP$, the optimal data departure curve can be obtained as the tightest string whose ends are tied to the origin and  the point $(T, A(T,\cachingP))$, which is represented in Fig. \ref{fig_probRepresentation} with the dashed lines. 
In particular, if the instantaneous cost, $g(\cdot)$, is strictly convex, then this 
is the unique optimal data departure curve;
contrarily, if $g(\cdot)$ is linear multiple optimal departure curves exist.
\end{lemma}

The free memory space in the cache can be obtained as  $B(t,\cachingP) - \departure$, $\forall t$.
Focusing on Policy 1
 (see Fig. \ref{fig_probRepresentation}(a)),
  the cache is full at  $t=\Ts$, and all the data in the cache belongs to \file[2] and/or \file[4], which have been pre-downloaded to equalize the rates in the first and second time slots.
As for  Policy 2 (see Fig. \ref{fig_probRepresentation}(b)), the cache is full at  $t=2 \Ts$, and  exclusively contains \file[2].
Note that  by caching $\file[2]$ in the second slot the upper bound is tightened (the net cache capacity is reduced) while the lower bound is relaxed (the demand at the third slot is reduced).

From the previous discussion, 
two questions arise: 
i) ``which of the two caching policies achieves the lowest MBS cost?'',
and ii) ``is any of these policies the optimal one?''.
One might be tempted to think that the caching policy $\vec{\tilde c}$ has a lower cost since fewer data has to be transmitted; however, this does not necessarily hold true since the caching policy $\vec{\hat c}$ might achieve a lower cost by equalizing the rate across time slots. 
In practice, the jointly optimal  transmission and local caching policies must be obtained by solving \eqref{eq:problem}, which turns out to be challenging since this problem belongs to the class of infinite-dimensional optimization problems \cite{infiniteDimensionalOptimization_fattorini1999}.

\vspace{-1em}
\subsection{Jointly optimal caching and transmission policies}
\label{sec_FormulationLinear}
To solve the infinite-dimensional problem in \eqref{eq:problem}, 
we first derive some structural properties of the  optimal strategy.
Then, leveraging on these properties, we will formulate  \eqref{eq:problem} as a finite-dimensional convex program of affordable complexity.
As shown next, the optimization variables of the resulting problem are the amount of data to be cached in each slot  for each request, \cached, $\forall n,u,$ and the transmission rate of the MBS at each slot, \rateEpoch, $\forall n$.

As illustrated in Fig. \ref{fig_probRepresentation}, the caching policy changes the shape of the upper and lower bounds on the data departure curve.
 For example, in Fig. \ref{fig_probRepresentation}(b), we have observed that the data locally cached in the second slot reduces the demand in the third slot.
Since the caching rate can have continuous variations over time, we can potentially have arbitrary non-decreasing curves as the upper and lower bounds, $B(t,\cachingP)$ and $A(t,\cachingP)$.
However, these curves are coupled through the caching policy $\cachingP$.
 In other words, 
the caching rate of a certain  request determines the reduction in the demand rate of the subsequent request.
The following lemma shows that (within a time slot) caching data at a constant rate turns out to be optimal.
\begin{lemma}[Constant rate caching is optimal]
\label{lemma:constantCachingSBS}
The (not necessarily  unique) optimal local caching rate is a step-wise function that can be written as
$\optimalCaching[(t)] = (\optimalCachingUser)_{u=1}^{U}$, where
$\optimalCachingUser = \sum_{n=1}^{N}  (\cached^\star/\Ts) \operatorname{rect} ( (t - (n-1/2)\Ts )/\Ts)$,
and $\cached^\star$ denotes the optimal amount of cached data for the  request of the $u$-th user at slot $n$.
\end{lemma}
\begin{IEEEproof}
See the Appendix.
\end{IEEEproof}

Since $\servingUser(t)$ and  $\optimalCachingUser$  are step-wise functions whose value can only change at slot transitions, we know that $A(t,\optimalCaching[])$ and $B(t,\optimalCaching[])$ are piece-wise linear functions (c.f. Definitions \ref{def:maxD} and \ref{def:minD}) whose slopes can only change at slot transitions. 
Consequently, we can obtain the following properties of the optimal transmission strategy.
\begin{lemma}
\label{lemma:departure}
The (not necessarily  unique) optimal data departure curve, \optimalDeparture, can be written as a piece-wise linear function, whose  rate (or, equivalently, the slope of \optimalDeparture)
may only change at time instants $n\cdot\Ts$, $n=1,\dots, N-1$\ie
 $r^\star(t) =  \sum_{n=1}^{N}  \rateEpoch^\star  \operatorname{rect} ((t - (n -1/2)\Ts )/\Ts)$, where  $\rateEpoch^\star$  denotes the optimal transmission rate of the \ac{MBS} at the $n$-th slot.
Additionally,  if the rate increases at the $n$-th slot transition ($ \rateEpoch^\star <  \rateEpoch[n+1]^\star $), then $\optimalDeparture[{n\Ts}, r^{\star}] = B(n\Ts, \optimalCaching[])$; and if
the rate decreases at the $n$-th slot transition ($ \rateEpoch^\star >  \rateEpoch[n+1]^\star $), then $\optimalDeparture[{n\Ts}, r^{\star}] = A(n\Ts, \optimalCaching[])$.
\end{lemma}
\begin{IEEEproof}
The proof follows similarly to \cite[Lemmas 5 and 6]{Gregori_QoS_2012} by
identifying $n\Ts$ as $\ell_m$,  $B(n\Ts, \optimalCaching[])$ as $D_{max}^{(m)}(\ell_m)$, and $A(n\Ts, \optimalCaching[])$ as $D_{min}^{(m)}(\ell_m)$.
\end{IEEEproof}

From Lemmas \ref{lemma:constantCachingSBS} and \ref{lemma:departure},
we can equivalently rewrite the original problem in \eqref{eq:problem} as a function of the MBS rates at each slot,
$\rate \triangleq (\rateEpoch)_{n=1}^{N}$, and 
cached data units  at the SBS for each user request and time slot,
$\cachedAll\triangleq ( (\cached)_{u=1}^{U})_{n=1}^{N}$:
\begin{subequations}
\label{eq:problemLinear}
\begin{align}
\minimize_{\rate, \cachedAll}  & \quad \sum_{n=1}^N \Ts g(\rateEpoch)& \\
\st 
& \quad\sum_{\ell =1}^n \Ts \rateEpoch[\ell] \leq C + \sum_{\ell =1}^n \sum_{u=1}^{U}  \Ts \servingNU[\ell u] - \cached[\ell u]  ,& \forall n, \label{eq:problemLinearC1} \\
& \quad  \sum_{\ell =1}^n \Ts \rateEpoch[\ell] \geq 
\sum_{\ell =1}^n \sum_{u=1}^{U} \Ts \servingNU[\ell u] -  \cached[\bar{\rhoFunction}(\ell,u)] ,& \forall n, \label{eq:problemLinearC2} \\
 &\quad \rateEpoch \geq 0,& \forall n, \\
&\quad 0 \leq \cached \leq \Ts \servingNU,& \forall n,u, \label{eq:problemLinearC4}
\end{align}
\end{subequations}
where the constraints \eqref{eq:problemLinearC1}-\eqref{eq:problemLinearC4} correspond to the discrete versions of the constraints in \eqref{eq:c2}-\eqref{eq:c4}, respectively.
The function $(n',u') =\bar{\rhoFunction}(n,u)$
returns the slot, $n'$, and  user, $u'$, 
of the previous request of the file associated to $(n,u)$, or returns $(-1,-1)$ if it is the first request of the file. 
Note that $\bar \rhoFunction$ is the discrete version of the function $\rhoFunction$; and as before, we define $\cached[-1-1] \triangleq 0$.

\begin{remark}
In \eqref{eq:problemLinear}, 
we have 
considered  that the amount of cached  data, \cached, is a nonnegative real number. 
Note that if we introduce an integer constraint to enforce  data unit granularity (e.g., bit), then the problem  in \eqref{eq:problemLinear}  becomes an integer program with its inherent complexity.
In practice, as the data unit granularity (bit) is sufficiently small in comparison to the files sizes (of several Mbits) and cache capacity (of several Gbits), the integer constraint can be relaxed without jeopardizing the performance.
Consequently, \eqref{eq:problemLinear} is a convex program (since the objective function is convex and the constraints are affine), and, thus, can be solved efficiently.
\end{remark}

By studying the \acl{KKT} conditions of the primal problem in \eqref{eq:problemLinear}, it is 
difficult to derive the structure of the optimal solution $\{ \rate^{\star}, \cachedAll^{\star} \}$ due to the constraints in  \eqref{eq:problemLinearC1} and \eqref{eq:problemLinearC2} that couple the optimization variables.
However, the structure of the optimal primal variables can be obtained by resorting to dual decomposition.
From convex optimization theory \cite{boyd_convex_2004}, the solution of the dual problem,
$\max_{\vecs \lambda, \vecs \mu}\delta(\vecs \lambda, \vecs \mu)$,
provides a lower bound on the primal problem in \eqref{eq:problemLinear}.
We have defined $\vecs \lambda = (\lambda_{n})_{n=1}^{N}$ and $\vecs \mu = (\mu_{n})_{n=1}^{N}$, where $\lambda_{n}$ and $\mu_{n}$ are the Lagrange multipliers associated to the $n$-th cache capacity and demand constraints, respectively.
The function
$\delta(\vecs \lambda, \vecs \mu)$ stands for the dual function that is defined as follows \cite{boyd_convex_2004}:
\begin{eqnarray}
\label{eq:dualFunction}
\delta(\vecs \lambda, \vecs \mu) = \minimize_{\rate, \cachedAll }   &&  \hspace{-1em}  \mathcal{L}(\rate, \cachedAll, \vecs \lambda, \vecs \mu) \\
\st 
 && \rateEpoch \geq 0,  \forall n, \quad  0 \leq \cached \leq \Ts \servingNU, \forall n, u, \nonumber 
\end{eqnarray}
where $\mathcal{L}(\rate, \vec q, \vecs \lambda, \vecs \mu)$ denotes the Lagrangian\ie
 $\mathcal{L}(\rate, \vec q, \vecs \lambda, \vecs \mu)= \sum_{n=1}^N \Ts g(\rateEpoch) +  \lambda_{n} 
 \big( - C +  \sum_{\ell =1}^n \Ts \rateEpoch[\ell]  -  \sum_{u=1}^{U} \Ts \servingNU[\ell u] - \cached[\ell u] \big) \nonumber  -  \mu_{n} 
 \big( \sum_{\ell =1}^n \Ts \rateEpoch[\ell] -  \sum_{u=1}^{U} \Ts \servingNU[\ell u] -  \cached[\bar{\rhoFunction}(\ell,u)] 	\big).$

Since the primal problem in \eqref{eq:problemLinear} is convex and the Slater constraint qualification holds, the duality gap (difference between the optimal values of the primal and dual problems) is zero \cite{boyd_convex_2004}. 
To solve the dual problem, we have implemented the projected subgradient method,  presented in Algorithm \ref{Alg:projectedSubgradient}, that guarantees convergence
to the optimal dual variables, $\vecs \lambda^{\star}$ and $\vecs \mu^{\star}$,
 if the updating step size $\epsilon^{(k)}$ is correctly chosen \cite{bertsekas_convexOptimization_2003}.

 \begin{algorithm}[t]
\footnotesize
	\caption{Projected subgradient}
\label{Alg:projectedSubgradient}
\renewcommand{\algorithmicrequire}{\textbf{Initialization:}}

	\begin{algorithmic}
	
	\Require
	\State  Set $k:= 0$ and initialize $\vecs \lambda^{(0)}$  and $\vecs \mu^{(0)}$ to any value such that $\vecs \lambda^{(0)} \succeq \vec 0$, $\vecs \mu^{(0)} \succeq \vec 0$.

	
	\State \textbf{Step 1}: If a termination condition is met, the algorithm stops. 
	\State \textbf{Step 2}: Compute $\rate^{(k)} , \cachedAll^{(k)}$ as the solution to the problem in \eqref{eq:dualFunction} given the current multipliers, $\vecs \lambda^{(k)}$  and $\vecs \mu^{(k)}$.

	\State \textbf{Step 3}:	Update the dual variables following the  subgradient\ie
 $ [\vecs \lambda^{(k+1)}]_n = \lambda_n^{(k+1)}$ and $ [\vecs \mu^{(k+1)}]_n = \mu_n^{(k+1)}$, $\forall n$, with  
$$\lambda_n^{(k+1)} = \left[\lambda_n^{(k)} + \epsilon^{(k)}
							  \left( - C + \sum_{\ell =1}^n \Ts \rateEpoch[\ell]^{(k)}  - \sum_{u=1}^{U} \Ts \servingNU[\ell u] - \cached[\ell u]^{(k)} \right) 
					  \right]^+  $$ 
$$\mu_n^{(k+1)} = \left[\mu_n^{(k)} - \epsilon^{(k)}
							  \left(  \sum_{\ell =1}^n \Ts \rateEpoch[\ell]^{(k)} -  \sum_{u=1}^{U} \Ts \servingNU[\ell u] -  \cached[\bar{\rhoFunction}(\ell,u)]^{(k)} 	 \right) 
					  \right]^+  .$$
	\State \textbf{Step 4}: Set $k := k+1$ and go to Step 1.
	\end{algorithmic}
	\end{algorithm}

Step 2 of Algorithm \ref{Alg:projectedSubgradient}  requires to solve the problem in \eqref{eq:dualFunction}, where the optimization variables are $\rate, \vec q,$ and the Lagrange multipliers ($\vecs \lambda$  and $\vecs \mu$) are fixed. 
To do so, we first rewrite the Lagrangian by 
reordering the sums over $n$ and $\ell$, which allows us to
separate the terms  associated to each  \rateEpoch and \cached\ie
$\mathcal{L}(\rate, \vec q, \vecs \lambda, \vecs \mu)
 = \sum_{n=1}^N \Ts g(\rateEpoch) -  \rateEpoch \Ts \big( \sum_{\ell =n}^N  \mu_{\ell}-\lambda_{\ell } \big)
  + \sum_{n=1}^N \sum_{u=1}^U  \cached  \big(  \sum_{\ell =n}^N \lambda_{\ell } - \sum_{\ell =\psi(n,u)}^{N} \mu_{\ell} \big) +
  \sum_{\ell =1}^{N} \mu_{\ell} (\sum_{n =1}^\ell \sum_{u=1}^{U} \Ts \servingNU) -\sum_{\ell =1}^N \lambda_{\ell } ( C + \sum_{n =1}^\ell \sum_{u=1}^{U} \Ts \servingNU ).  $
The function $\psi(n,u)$ returns the slot index of the subsequent request of the file being served at slot $n$ to user $u$.
Now, the problem in \eqref{eq:dualFunction} is decoupled in the optimization variables ($ \rate$ and  $\vec q$) and 
can be easily solved by decomposing it into the following simpler subproblems:
\begin{align}
\label{eq:dualDecomp1}
 & \hspace{0.5cm} \minimize_{\rateEpoch \geq 0}   \quad  \Ts g(\rateEpoch) -  \rateEpoch \Ts \left(\sum_{\ell =n}^N  \mu_{\ell } -\lambda_{\ell } \right),   &\forall n,
\\
\label{eq:dualDecomp2}
& \minimize_{0 \leq \cached \leq \Ts \servingNU }   \quad  \cached  \left( \sum_{\ell =n}^N \lambda_{\ell }  -  \sum_{\ell =\psi(n,u)}^{N} \mu_{\ell} \right),  &\forall n,u.
\end{align}
Let
$\bar r_n$ be the solution to the equation
$\d g(\rateEpoch)/ \d\rateEpoch = \sum_{\ell =n}^{N} \mu_{\ell } -  \lambda_{\ell }$.
If $\bar r_n$ is real and positive, the optimal solution to \eqref{eq:dualDecomp1} is $\rateEpoch^{\star}(\vecs \lambda, \vecs \mu) = \bar r_n$;
otherwise,    it is $\rateEpoch^{\star}(\vecs \lambda, \vecs \mu) = 0$. 

\begin{corollary}
When the objective is the minimization of the energy consumption over the backhaul link ($g(\rateEpoch)= (\exp(\rateEpoch) - 1)/h$),
the optimal  solution to \eqref{eq:dualDecomp1} is found as
$\rateEpoch^{\star}(\vecs \lambda, \vecs \mu) = \log(h (\sum_{\ell =n}^{N} \mu_{\ell } -  \lambda_{\ell }) )$
 if $h (\sum_{\ell =n}^{N} \mu_{\ell } -  \lambda_{\ell }) > 1$ and
$\rateEpoch^{\star}(\vecs \lambda, \vecs \mu) = 0$, otherwise.
\end{corollary}

The solution to \eqref{eq:dualDecomp2} is
\begin{equation}
\label{eq:dualFunctionSolQ}
\cached^\star(\vecs \lambda, \vecs \mu) =
 \left\{ \,
 \begin{IEEEeqnarraybox}[][c]{l?s}
\IEEEstrut
 0 & if $W_{nu}>0$, \\
\Ts \servingNU & if $W_{nu}< 0$, \\
{\bar{q}_{nu} \in [0, \Ts \servingNU]} & if $W_{nu} = 0,$
\IEEEstrut
\end{IEEEeqnarraybox}
\right.
\quad
\end{equation}
with
$W_{nu} \triangleq  \sum_{\ell =n}^N \lambda_{\ell }  -  \sum_{\ell =\psi(n,u)}^{N} \mu_{\ell}.$
Accordingly, the primal variables at the $q$-th iteration of the subgradient, which are necessary in Step 2 of Algorithm \ref{Alg:projectedSubgradient}, are given by
$\rate^{(k)} = \big( \rateEpoch^{\star}(\vecs \lambda^{(k)}, \vecs \mu^{(k)}) \big)_{n=1}^{N}$ and $
 \cachedAll^{(k)}= \big(\big(\cached^\star(\vecs \lambda^{(k)}, \vecs \mu^{(k)})\big)_{u=1}^{U} \big)_{n=1}^{N} $, where $\vecs \lambda^{(k)}$ and $ \vecs \mu^{(k)}$ denote the Lagrange multipliers at the $q$-th iteration of the subgradient.
 
When the subgradient algorithm converges to the optimal Lagrange multipliers, $\{\vecs \lambda^\star, \vecs \mu^\star\}$,
the duality gap is zero\ie the optimal solution of the dual and primal problems are the same.
Note that  given the optimal dual variables, $\{\vecs \lambda^{\star}, \vecs \mu^{\star}\}$, there might be multiple minimizers of the problem in \eqref{eq:dualFunction}.
Precisely,  $\cached^\star(\vecs \lambda^{\star}, \vecs \mu^{\star})$ can take multiple values when $W_{nu} = 0$ (see \eqref{eq:dualFunctionSolQ}).
Then, the optimal primal variables $\{\rate^{\star}, \vec q^{\star}\}$ are within the set of minimizers of $\delta(\vecs \lambda^{\star}, \vecs \mu^{\star})$ in  \eqref{eq:dualFunction}; 
in particular, $\{\rate^{\star}, \vec q^{\star}\}$
 are the minimizers that are feasible  in the primal problem
\eqref{eq:problemLinear} and satisfy the slackness conditions \cite{bertsekas_convexOptimization_2003}.
In practice, to avoid waiting until the exact convergence to $\{\vecs \lambda^\star, \vecs \mu^\star\}$,
the average across iterations of the primal iterates can be used as an approximate solution to the problem in \eqref{eq:problemLinear} \cite{Nedic_Approximate_2009}.

Interestingly, it turns out that
the parameter $W_{nu}$, which only depends on the Lagrange multipliers, characterizes the caching policy:
if $W_{nu}$ is positive, the associated file is not cached;
while, if $W_{nu}$ is negative the file is completely cached;
and, finally, if $W_{nu} = 0$ the SBS caches a portion of the file (the exact amount of cached data units must be obtained as mentioned in the previous paragraph). 
Additionally, from the expression of $W_{nu}$, we observe that
the SBS caching policy prioritizes the
 files that are requested again in the near future.\footnote{
From the KKT optimality conditions, we have  
 $\mu_n > 0$  if the $n$-th  demand constraint in \eqref{eq:problemLinearC2} is satisfied with equality (and zero otherwise).
Consider that users $u$ and $u'$ request different files and that the subsequent request of these files appears first for user $u$\ie $\psi(n,u) < \psi(n,u')$.
Then, from the expression of $W_{nu}$ in \eqref{eq:dualFunctionSolQ}, we have  $W_{nu} \leq  W_{nu'}$  and, as a result,
the SBS prioritizes 
 caching the file requested by user $u$.
}


%

\vspace{-1em}
\section{Caching at user devices}
\label{sec:D2D}

\vspace{-0.5em}
\subsection{System model and problem formulation}
\label{sec:D2D_model}

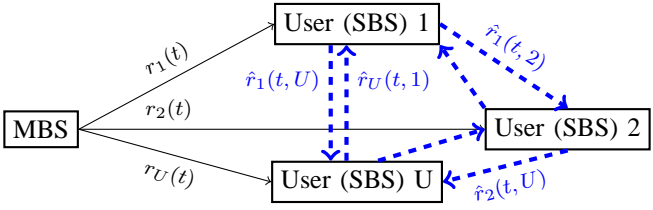
\begin{figure}
\centering
\begin{tikzpicture}[
	remember picture,
	 scale=0.7,
   	   important line/.style={thick},
    ]
    

	 \node[draw, thick, rectangle, text centered] (MBS) at (-5,0){ MBS };
 	 \node[draw, thick, rectangle, text centered] (user1) at (1,2){ User (SBS) 1};
  	 \node[draw, thick, rectangle, text centered] (user2) at (5,0){ User  (SBS) 2};
   	 \node[draw, thick, rectangle, text centered] (userU) at (1,-1){ User (SBS) U};
   	 
   	 \draw[->] (MBS.east)--(user1.west) node[midway,sloped, above, font = \footnotesize] {$\rateUser[1](t)$};
	 \draw[->] (MBS.east)--(user2.west) node[midway,sloped, above, xshift=-1.5cm, font = \footnotesize] {$\rateUser[2](t)$};
 	 \draw[->] (MBS.east)--(userU.west) node[midway,sloped, below, font = \footnotesize] {$\rateUser[U](t)$};
 	 
   	 \draw[->,dashed, blue, ultra thick] (user1.east)--(user2.north) node[midway,sloped, above, font = \footnotesize] {$\rateUserUser[t,2]{1}$};
  	 \draw[->,dashed, blue, ultra thick] ([xshift = -0.5cm]user1.south)--([xshift = -0.5cm]userU.north) node[midway, left, font = \footnotesize, yshift=0.3cm] {$\rateUserUser[t,U]{1}$};

   	 \draw[->,dashed, blue, ultra thick] (user2.north west) -- (user1.south east);
	 	 \draw[->,dashed, blue, ultra thick] (user2.south) -- (userU.east) node[midway, below, sloped, font = \footnotesize] {$\rateUserUser[t,U]{2}$};;

  	 \draw[->,dashed, blue, ultra thick] ([xshift = -0.2cm]userU.north) -- ([xshift = -0.2cm]user1.south)node[midway, right, font = \footnotesize, yshift=0.3cm] {$\rateUserUser[t,1]{U}$};
   	 \draw[->,dashed, blue, ultra thick] ([xshift = 0.4cm]userU.north) -- (user2.west);
  	 
%


\end{tikzpicture}
\caption{System model when caching is performed at
the user devices, which act as an SBS for the other users through D2D communications.}
\label{fig:systemModelMBSD2D}
\end{figure} 
In this section, as depicted in Fig. \ref{fig:systemModelMBSD2D},
we consider 
a region in space covered by an MBS that must serve the demand of $U$ users.
The  MBS allocates an \textit{orthogonal} channel to each user, whose transmission rate is denoted by $\rateUser(t)$, $u= 1, \dots,U$.
We assume that users 
cooperate with the network operator (possibly in exchange of  incentives) by 
acting as an SBS for the rest of the users through dedicated D2D links.
We consider that
users
are closely located; thus, a certain user $u$ can act as an SBS for any other user $u' \neq u$.
The rate in the D2D link from SBS (user) $u$ to user $u'$ at time $t$ is denoted by  $\rateUserUser{u}$, $u' \neq u = 1, \dots,U$.
We assume that the D2D links operate over different frequency resources than those used by the MBS.
A  user is allowed to download only its own traffic from the MBS;
that is, users do not download content that is of no interest to them, solely to serve another user.
However,  downloaded files can be locally cached to later serve other users through the D2D links.

\begin{figure}
\centering
\begin{tikzpicture}[
	scale=0.56,
   	   important line/.style={thick},
   	   every node/.style={font=\small}
     ]


		
			\coordinate (Decoding) at (-6,0);
			\coordinate (Coding) at (8,0);
			\draw[very thick] (-6,-5) rectangle (2.5,2.4);
			\draw[dashed, thick] (-4.7,-2) rectangle (0.7,1.7);
			\node[ font = \scriptsize] at (-2,2){\textbf{Cache of capacity $C_u$}};

			\node[align=center,  below, text width= 2.5cm] at (-2,1.5) (preBuffer)    {Pre-downloading buffer};
		    \draw[very thick](preBuffer.north west)--(preBuffer.north east)--(preBuffer.south east)--(preBuffer.south west);
	
			\node[draw,circle, inner sep=0.10pt] (plus) at ([xshift= 1cm]preBuffer.east){+};
			\node[draw,circle, inner sep=0.10pt] (plus2) at ([xshift= 2cm]plus){+};
			\draw (preBuffer.east)-- (plus.west);
			\draw[draw,  thick]   ([xshift= 0.5cm, yshift= 0.2cm] preBuffer.east)  -- ([xshift= 0.5cm, yshift= -0.2cm] preBuffer.east)   node[below]{$\alpha$};
			
			\draw[->, thick] ([xshift=-0.5cm]preBuffer.west)-- (preBuffer.west) node[midway, right,xshift=-.2cm, yshift=1.25cm]{$\rateUser(t)$};
			
			\node[draw, rectangle,align=center,  text width= 1.5cm, very thick] at ([yshift=2.5cm] plus2.north)(user){\textbf{User Module}};
			\draw[->,blue, ultra thick]  ([yshift=0.5cm] user.north) node[above, black]{SBSs}-- (user.north);
			\draw[->, solid, blue, ultra thick] (user.south) -- (plus2.north) node[midway, right]{$\rateUserUser[t,u]{u'}, \forall u' \neq u$} ;

			\draw[-,thick]  ([yshift=3cm, xshift=-0.5cm] preBuffer.west) node[above]{MBS} -- ([xshift=-0.5cm] preBuffer.west) ;
						
			\node[align=center,  below, text width= 2.5cm, thick] at (-2,-.25) (locBuffer)    {Local caching buffer};
		    \draw[very thick] (locBuffer.north west)--(locBuffer.north east)--(locBuffer.south east)--(locBuffer.south west);			
			
			\node (demand) at ([xshift=4cm]plus2.east)  {Aplication};
			\draw[->]  (plus.east)--(plus2.west) (plus2.east)--(demand.west) node[below, text width = 1.5cm]{$\demandUser = d_{u}(t,u)$};
			\draw[->, very thick, red] ([xshift=0.5cm]plus2.east)--++(0,-4.25)-- ++(-6,0)node[below, midway]{$\cachingDtoDUU[t,u]{u}$} -- ([xshift=-1cm,yshift=-2.5cm]locBuffer.west) |-([yshift=0.4cm]locBuffer.west) ;

			\draw[->] 	([yshift = +0.4cm ] locBuffer.east)-|(plus.south) node[yshift= -0.2cm, xshift= -0.2cm,midway, right]{$\ell_{u}(t)$};			
			\draw[->, dashed, blue, ultra thick] 	([yshift = -0.4cm ] locBuffer.east)-- ( [yshift = -0.4cm ] locBuffer.east -| Coding.south) node[midway, xshift= 2cm, below, text width=3cm]{$d_u(t,u')= \rateUserUser{u}, \forall u'\neq u$} node[right] {D2D} ;
			\draw[->, very thick, red] 	([xshift=3cm,yshift = -0.4cm ]locBuffer.east)-- ++ (0,-1.75) -- ([xshift=-0.5cm, yshift=-2.15cm]locBuffer.west) node[midway, above]{$\cachingDtoDUU[t,u']{u}, \forall u'\neq u$}|-([yshift=-0.25cm] locBuffer.west);	
				
			\node[above] at (-1.75,-5) {\textbf{SBS Module}};

\end{tikzpicture}
\caption{Block diagram of the $u$-th user terminal.
 The solid lines correspond to data streams associated to traffic of the $u$-th user, and the dashed ones represent traffic served to other users, $u'\neq u$, that is transmitted through the D2D links.}
\label{fig:userTerminal}

\end{figure}
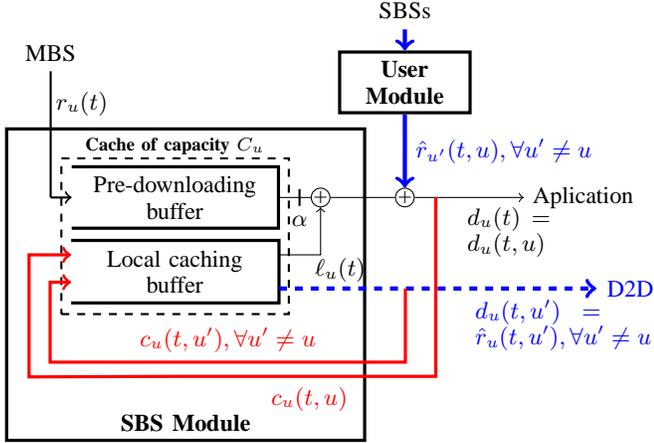

As represented in Fig. \ref{fig:userTerminal}, the D2D user terminals are composed of two main modules: the user module and the SBS module.
The \textit{user module} acts exactly as a user terminal in the previous scenario\ie it receives data from 
the SBSs (now from the SBS modules of other users) and feeds it directly to the application layer.
For fair comparison with the previous scenario, 
we do not allow  data to be cached within the user module. 
As a result,
if user $u$ caches data at time $t$ to reduce the demand from the MBS of another user $u'$ at time $t'$, $t' > t$,
then user $u$ must send this data over the D2D link at time $t'$, and not earlier.


The \textit{SBS module} at the user terminal essentially acts as the SBS terminal in the previous scenario;
the main difference is that the  SBS module in the $u$-th user terminal 
is  allowed to download only contents corresponding to its own demand  from the MBS\ie  $d_u(t,u) \triangleq d_u(t)$.
This module contains a cache memory of capacity $C_{u}$,  represented with two different virtual buffers to ease interpretation. 
The first virtual buffer is used to represent pre-downloaded
contents from the MBS associated to 
the $u$-th user's demand.
The second virtual buffer represents locally cached data from previous demands.
When the $u$-th user serves its demand, at a rate $d_u(t,u)$, to the application layer,
data can be cached at a rate $\cachingDtoDUU[t,u]{u} \leq d_u(t,u)$.
This cached data can be later requested by other users;
in particular, $d_u(t,u')$ denotes the demand  of	 user $u$ from user $u'$ at time $t$, which is served through the D2D links; accordingly, we have $d_u(t,u')= \rateUserUser{u}$. 
The caching rate associated to demand $d_u(t,u')$ is denoted by $\cachingDtoDUU[t,u']{u}$.
Finally, $\ell_u(t)$ denotes the rate at which data is removed from the local cache of user $u$ to reduce its own demand, which is used if a file is requested twice by the user.
To simplify the notation, in the remainder of the paper we refer to variable $\ell_u(t)$ as  $\rateUserUser[t,u]{u}$. However, note that this data stream does not require D2D resources.

As before, the aim is to jointly design the transmission  and  caching rates,
$\rateMBS \triangleq  (\rateUser{(t)})_{u=1}^{U}$, $\ratesDtoD(t)\triangleq  ((\rateUserUser[t,u']{u})_{u'=1}^{U})_{u=1}^{U}$, and $\caching \triangleq  ((\cachingDtoDUU[t,u']{u})_{u'=1}^{U})_{u=1}^{U}$, that minimize a general cost function  on the rates of the MBS,  $\rateMBS$, and D2D links, $\ratesDtoD(t)$.
This cost function accounts for: 
(i) the cost of transmissions from the MBS to different users, 
$\sum_{u=1}^U g_{u} (\rateUser(t)) $,  with $g_{u} (\rateUser(t))$ standing for the instantaneous cost of the $u$-th link;
and (ii) the cost of transmissions over the D2D links, $\sum_{u=1}^{U}\sum_{u'\neq u}\hat{g}_{u u'}( \rateUserUser[t,u']{u})$, where $\hat{g}_{u u'}( \rateUserUser[t,u']{u})$ denotes the cost of the D2D link from user $u$ to user $u'$.
Again, we assume that the functions  $g_{u}(\cdot)$,   $\hat{g}_{u u'}(\cdot)$, $\forall u, u'\neq u$, are 
 time invariant, convex, increasing, continuously differentiable, and
$g_{u}(0) = 0$, $\hat{g}_{u u'}(0) = 0$, $\forall u,  u'\neq u$.\footnote{
Notice that by
considering a different cost function for each link, it is possible to model, among others, different channel gains.}
As in Section \ref{sec:SBS}, different objective functions can be modeled by appropriately selecting the cost functions $g_{u}(\cdot)$  and  $\hat{g}_{u u'}(\cdot)$ (e.g., energy consumption, energy cost, bandwidth, or traffic minimization). 
As mentioned earlier, the operator must give incentives to users that transmit over the D2D links \cite{PengLi_incentiveD2D}. In this context, 
 $\hat{g}_{u u'}(\cdot)$ can represent the economical incentive, $ \xi_{U}\geq 0$, paid by the operator  to the users for the data transmitted  over the D2D links\ie   $\hat{g}_{u u'}(\rateUserUser[t,u']{u}) =  \xi_{U} \rateUserUser[t,u']{u}$.  The total incentive of user $u$ is $\int_{t=0}^{T} \sum_{u'\neq u}  \xi_{U}  \rateUserUser[t,u']{u}$. The total economical cost of the operator is the
 sum of the cost of energy used by the MBS, 
 $\int_{0}^{T} \xi_{MBS}\left (\sum_{u=1}^U W/U(\exp (\rateUser(\tau) U / W ) -1)\right ) \d \tau$, and the incentives paid to the users, $\int_{0}^{T} \xi_{U}\sum_{u=1}^U \sum_{u'\neq u}  \rateUserUser[\tau,u']{u}  \d \tau $. 
Observe that, if this cost function is adopted, no channel state information of the D2D links is required at the MBS.

For  problem tractability, we forbid users to simultaneously cache the same data by including the constraint 
$ \sum_{u'=1}^{U}\cachingDtoDUU[t,u]{u'} \leq  \demandUser, \quad \forall u.$
The implications of this assumption are later discussed in Remark \ref{optimalityD2D}.

Since we have one dedicated MBS link per user,
we need a 
data departure curve from the MBS to each user $u$\ie
$\departureUser = \int_{0}^{t} \rateUser(\tau) \d\tau.$
The $u$-th data departure curve is constrained from above by the $u$-th user cache capacity,  $C_u$ (through the maximum data departure curve $B_{u}$),  and from below by its net demand from the MBS (through the minimum data departure curve $A_{u}$). 

Next, we derive the expression of the minimum data departure curve for the 
transmission from the MBS to the  $u$-th user, to fulfill the user's demand \demandUser.
Since from the previous assumption the users' cache contents are non-overlapping, the net demand of user $u$ from the MBS (i.e., the demand in point $\alpha$ in Fig. \ref{fig:userTerminal}) is 
obtained by subtracting from \demandUser the sum of the rates in the D2D links to user $u$ and the locally cached data at user $u$, $\ell_u(t)= \rateUserUser[\tau,u]{u} $.
Thus, the lower bound on \departureUser reads as
$
 A_{u}(t, \ratesDtoD)
 \triangleq \int_{0}^{t} \demandUser[\tau] -  \sum_{\forall u'}\rateUserUser[\tau,u]{u'}  \d \tau.
$
The maximum data departure curve at user $u$,  $B_{u}(t, \cachingP, \ratesDtoD)$, 
can be obtained by adding $C_u$ to the minimum data departure curve,  $A_{u}(t, \ratesDtoD)$, and subtracting the locally cached data\ie 
$ B_{u}(t, \cachingP, \ratesDtoD ) \triangleq  C_{u} +  A_{u}(t, \ratesDtoD)
- \int_{0}^{t} \sum_{\forall u'}( \cachingDtoDUU[\tau,u']{u} -\rateUserUser[\tau,u']{u})  \d \tau, \nonumber $
where
the locally cached data is computed as the integral up to time $t$ of the difference between the data rates entering and leaving the local caching buffer.

Bearing all the above in mind, the problem 
with  optimization variables $\{\rateMBS,\ratesDtoD(t), \caching\}_{t\in[0,T]}$
is mathematically formulated as follows:
\begin{subequations}
\label{eq:problemD2D}
\begin{align}
\minimize &  \int_{0}^{T} \sum_{u=1}^U \left ( g_{u} (\rateUser(\tau)) +\sum_{u'\neq u} \hat{g}_{u u'}( \rateUserUser[\tau,u']{u}) \right ) \d \tau \\
\st \:\:\:
   &\departureUser \leq B_{u}(t, \cachingP, \ratesDtoD), \hspace{1.6cm} \forall u,  t \in [0,T], \label{eq:c2D2D} \\
  &\departureUser \geq A_u(t, \ratesDtoD), \hspace{2cm} \forall u, t \in [0,T],  \label{eq:c1D2D}\\
  &  \sum_{u'=1}^{U}\cachingDtoDUU[t,u]{u'} \leq  \demandUser,  \hspace{1.8cm} \forall u, t \in [0,T], \label{eq:c4D2D} \\
  & \cachingDtoDUU[t,u']{u} \leq  \cachingDtoDUU[\rhoFunction(t,u')]{u}, \hspace{0.5cm}   \forall u, u' \neq u, t \in [0,T], \label{eq:c5D2D}\\
   &\rateUserUser[t,u']{u} \leq  \cachingDtoDUU[\rhoFunction(t,u')]{u},  \hspace{1.2cm} \forall u, u', t \in [0,T], \label{eq:c6D2D}\\
  & \cachingDtoDUU[t,u']{u} \geq 0,  \rateUser(t)  \geq 0,  \rateUserUser[t,u']{u} \geq 0,  \forall u, u', t \in [0,T], \label{eq:c3D2D} 
\end{align}
\end{subequations}
where the constraints in \eqref{eq:c2D2D} prevent cache overflows at the users, and those in \eqref{eq:c1D2D} impose the fulfillment of the users' demands. 
The constraints in \eqref{eq:c4D2D} impose that the same data cannot be simultaneously cached by different users,
and those in \eqref{eq:c5D2D} and \eqref{eq:c6D2D} restrict  
the maximum caching rate and D2D transmission rate at user $u$ when serving the requests of other users $u' \neq u$ to the
rate locally available in the cache of user $u$, $\cachingDtoDUU[\rhoFunction(t, u')]{u}$, respectively.
The function $\rhoFunction(t,u)$ is defined in the previous section.\footnote{
The function $\rhoFunction(t,u)$ as defined in the previous section is 
inherently enforcing that if two users $u$ and $u'$ request the same file at a given time slot,
then both users have to download the data that is not cached at other users from the MBS (i.e, they cannot help  each other in the current slot for this file).
However, we could redefine the function $\rhoFunction(t,u)$  to allow instantaneous transmissions over the D2D links, 
by making the request of one user to point the other one\ie $\rhoFunction(t,u') = [t,u]$;
thus, only user $u$ downloads the data from the MBS, which is instantaneously sent to user $u'$ through the D2D link.
}
Finally, the constraints in  \eqref{eq:c3D2D} 
impose nonnegative caching and transmission rates, respectively.
\begin{remark}
\label{optimalityD2D}
For problem tractability, we have included the constraint \eqref{eq:c4D2D} of caching non-overlapping data at the users.
This assumption is without loss of optimality when the D2D link costs are equal and linear\ie  $\hat{g}_{u u'}( x) = \xi_{U} x, \forall u, u'\ne u$, which, as mentioned earlier, characterizes the incentives paid by the operator to users. 
Then, there exists an optimal solution where users do not cache overlapping contents.
Further gains can be achieved by caching overlapping contents in the general case of unequal or non-linear D2D cost functions.
However, due to the corresponding combinatorial structure, the optimal solution considering overlapping caching at users is elusive, and is left as an open problem for future work. 
\end{remark}

\vspace{-0.5em}
\subsection{Jointly optimal strategy}
\label{sec_FormulationLinearD2D}
As in Section \ref{sec:SBS}, we
first derive some structural properties of the caching policy $\cachingP$, the  D2D transmission policy \ratesDtoD, and the data departure curves, $\departureUser$,
which allow us
to reformulate the problem with a finite number of optimization variables.
In the following lemma
 we show that, within each slot, it is optimal that the users cache and transmit data at constant rate.
\begin{lemma}
\label{lemma:constantCachingD2D}
The  (not necessarily unique) optimal  caching rate and D2D transmission rates in problem \eqref{eq:problemD2D} can be written as a piece-wise constant functions
$\optimalCachingUser[t,u'] = \sum_{n=1}^{N} (\cachedOptimal[u]{n,u'}/\Ts) \operatorname{rect} ( (t - (n-1/2)\Ts )/\Ts)$ and 
$\rateUserUser[t,u']{u} = \sum_{n=1}^{N} (\rateUserUserOptimal[u]{n,u'}/\Ts) \operatorname{rect} ( (t - (n-1/2)\Ts )/\Ts)$, where
 $\cachedOptimal[u]{n,u'}$ is the optimal amount of cached data at user $u$ for the  request of user $u'$ at slot $n$, and 
 $\rateUserUserOptimal[u]{n,u'}$ is the optimal amount of transmitted data from user $u$ to user $u'$  within slot $n$.
\end{lemma}
\begin{IEEEproof}
The proof follows similarly to the proof of Lemma \ref{lemma:constantCachingSBS} and is omitted for brevity.
\end{IEEEproof}

Since the optimal local caching rate and D2D transmission rates are piece-wise constant, we know that the constraints in \eqref{eq:c2D2D} and \eqref{eq:c1D2D} are piece-wise linear, and the slopes of the constraints can only change at some slot transition.
From this, and similarly to Lemma \ref{lemma:departure}, we can prove that the optimal data departure curve for each user 
can be written as a piece-wise linear function. Thus, the associated transmission rate from the MBS to each user reads as
$\rateUser^\star(t) =  \sum_{n=1}^{N}  \rateEpoch[nu]^\star  \operatorname{rect} (t - (n -1/2)\Ts /\Ts)$, where  $\rateEpoch[nu]^\star$  is the optimal transmission rate from the \ac{MBS} to user $u$ at the $n$-th slot.

%
The problem in \eqref{eq:problemD2D} can be equivalently rewritten 
in terms of cached data $\cachedDtoD{n,u'}$,
the transmitted data at the D2D links  $\rateUserUserDiscrete[u]{n,u'}$,
 and the transmission rates from the MBS to each user at each slot, \rateEpoch[nu], as
\begin{subequations}
\label{eq:problemLinearD2D}
\begin{align}
\minimize  & \sum_{u=1}^U  \sum_{n=1}^N \Ts \left (g_{u}(\rateEpoch[nu] ) 
 +  \sum_{u'\neq u} \hat{g}_{uu'}\left (\frac{\rateUserUserDiscrete[u]{n,u'}}{\Ts}\right ) \right )\\
\st 
&  \sum_{\ell =1}^n \Ts \rateEpoch[\ell u]  \leq C_u +  \sum_{\ell =1}^n \Big( \demandNU[\ell u] \Ts + \nonumber  \\
& \hspace{0.5cm} \sum_{u' = 1}^{U} \rateUserUserDiscrete[u]{\ell,u'} - \rateUserUserDiscrete[u']{\ell,u}- \cachedDtoD{\ell,u'} \Big), \hspace{0.2cm} \forall u, n,  \label{eq:problemLinearD2D1} \\
&   \sum_{\ell =1}^n \Ts \rateEpoch[\ell u] \geq  \sum_{\ell =1}^n \demandNU[\ell u] \Ts - \sum_{u'=1}^{U} \rateUserUserDiscrete[u']{\ell,u}, \hspace{0.5cm} \forall u,n, \\
& \sum_{u'=1}^U \cachedDtoD[u']{n,u} \leq \Ts \demandNU,\hspace{2.8cm} \forall n, u, \\
& \cachedDtoD[u]{n,u'} \leq \cachedDtoD[u]{\bar \rhoFunction(n,u')},\hspace{1.5cm} \forall n, u, u'\neq u, \\
& \rateUserUserDiscrete[u]{n,u'} \leq \cachedDtoD[u]{\bar \rhoFunction(n,u')},\hspace{2.2cm} \forall n, u, u', \\
& \cachedDtoD[u]{n,u'} \geq 0, \rateUserUserDiscrete[u]{n,u'} \geq 0, \rateEpoch \geq 0,\hspace{0.5cm}  \forall n, u, u',  \label{eq:problemLinearD2D2}
\end{align}
\end{subequations}
where the constraints in \eqref{eq:problemLinearD2D1}-\eqref{eq:problemLinearD2D2} stand for the discrete versions of the constraints in \eqref{eq:c2D2D}-\eqref{eq:c3D2D}, respectively. The function $\bar \rhoFunction$ is defined as in Section \ref{sec_FormulationLinear}.

The problem in \eqref{eq:problemLinearD2D} is a convex program because the objective function is convex and the constraints are linear.
Accordingly, it can be efficiently solved by\eg interior point methods.


\section{Numerical results}
\label{sec_Results}

In this section, we assess the performance of the proposed caching and transmission policies in both scenarios, namely, the \textit{SBS scenario} in Section \ref{sec:SBS} and the  \textit{D2D scenario} in Section \ref{sec:D2D}. 
We consider  $N= 20$ time slots of duration $10$ seconds each, and $F = 2000$ video files.
The file lengths are uniformly distributed in the interval $[0.3, 150] \ \mathrm{M nats}$ with mean file length $\mathbb{E}[\ell_j]= 75.15 \ \mathrm{M nats}$.
We assume that the probability of requesting file \file, $\theta_j$, is independent and identically distributed across time slots and users, and follows the Zipf distribution\ie 
$\theta_j = j^{-\gamma} /(\sum_{q=1}^{|\fileAlph| } q^{-\gamma}  )$.
Parameter $\gamma$ models the skewness of the file popularity;
when $\gamma= 0$, popularity is uniform and it becomes more skewed as $\gamma$ grows \cite{Blasco_ISIT_2014}.
We consider the Shannon power-rate function $g(r(t)) = W(\exp (r(t) / W ) -1)$, where $W$ is the channel bandwidth.
In the SBS scenario, the MBS allocates the whole bandwidth $W= 10 ~\mathrm{MHz}$ to communicate with the SBS; 
while in the D2D scenario,  the MBS splits evenly the total bandwidth across the $U$ users; as a result the bandwidth in each subchannel is $W= 10/U ~\mathrm{MHz}$.
The cache capacity $C$ is alternatively expressed by means of the percentage over the average requested data per user\ie  $\hat C =  (100 C) / ( N  \mathbb{E}\{\ell_j\})$.
For the D2D scenario, the total cache capacity is evenly distributed across users, $C_u = C/U$.
Unless otherwise stated, we set the number of users to $U=3$, the  Zipf distribution parameter to  $\gamma=1$, and the total cache capacity 
to  $\hat C=10$ ($C = 15.03  \mathrm{M nats}$).

We compare the proposed jointly optimal  transmission and caching strategies, obtained by solving \eqref{eq:problemLinear} and \eqref{eq:problemLinearD2D}, with
four sub-optimal strategies:
the \textit{No caching} strategy that serves as a benchmark for comparison with traditional systems without cache memory;
the \textit{\ac{LRU}}  caching algorithm that always keeps in the cache the most recently requested files \cite{Rizzo_CacheReplacement_2000};
the \textit{\ac{PDCA}} that  uses the cache only to pre-download data (see Policy 1 in Section \ref{sec:sysModelFormulation} and Fig. \ref{fig_probRepresentation}(a)) \cite{Gungor_proactiveCaching_2014};
and the \textit{\ac{LCA}} that  exploits only the local caching gain (i.e, the MBS transmits at the net demand rate without allowing pre-downloading).
The optimal policy, the PDCA, and the LCA are offline policies as non-causal knowledge of the file demand is required;
while the \textit{No caching}  and \textit{\ac{LRU}} strategies are online policies that may depend only on the previous file requests.

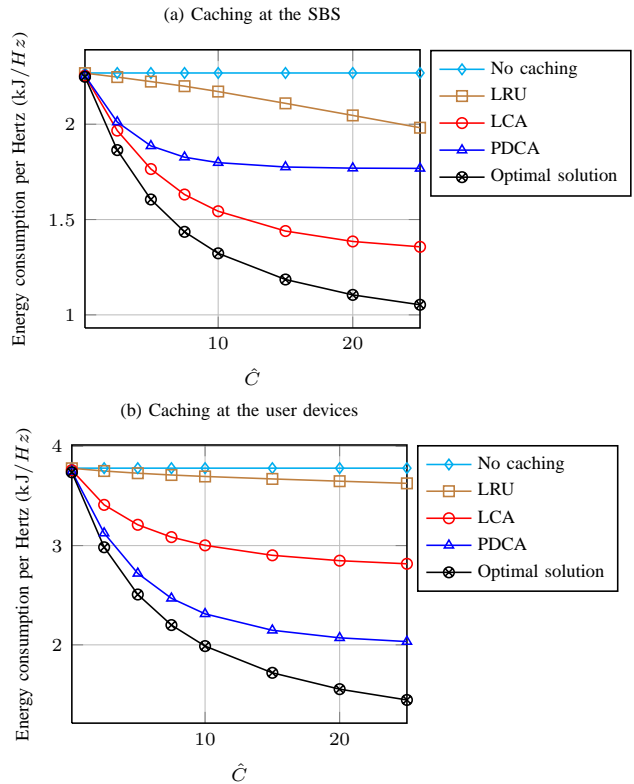
\begin{figure}
\centering
\begin{tikzpicture}
\begin{axis}[
scale=0.65,
grid,
ylabel={Energy consumption per Hertz ($\mathrm{kJ}/Hz$)},
xlabel={ $\hat C$ },
title= (a) Caching at the SBS,
legend pos=outer north east
]

\addplot  table[x=C, y expr=\thisrow{noCache}/1000]{file1.dat};
\addplot  table[x=C, y expr=\thisrow{LFU}/1000]{file1.dat};
\addplot  table[x=C, y expr=\thisrow{LC}/1000]{file1.dat};
\addplot  table[x=C, y expr=\thisrow{PD}/1000]{file1.dat};
\addplot  table[x=C, y expr=\thisrow{Optimal}/1000]{file1.dat};

\addlegendentry{No caching};
\addlegendentry{LRU};
\addlegendentry{LCA};
\addlegendentry{PDCA};
\addlegendentry{Optimal solution};
\end{axis}
\end{tikzpicture}
\begin{tikzpicture}
\begin{axis}[
scale=0.65,
grid,
ylabel={Energy consumption per Hertz ($\mathrm{kJ}/Hz$)},
xlabel={ $\hat C$ },
title= (b) Caching at the user devices,
legend pos=outer north east
]

\addplot  table[x=C, y expr=\thisrow{noCache}/1000]{file2.dat};
\addplot  table[x=C, y expr=\thisrow{LFU}/1000]{file2.dat};
\addplot  table[x=C, y expr=\thisrow{LC}/1000]{file2.dat};
\addplot  table[x=C, y expr=\thisrow{PD}/1000]{file2.dat};
\addplot  table[x=C, y expr=\thisrow{Optimal}/1000]{file2.dat};

\addlegendentry{No caching};
\addlegendentry{LRU};
\addlegendentry{LCA};
\addlegendentry{PDCA};
\addlegendentry{Optimal solution};
\end{axis}
\end{tikzpicture}
\vspace{-1em}
\caption{ Energy consumption of the \ac{MBS} with respect to  cache capacity ($\gamma = 1$, $U =3$). 
In (a), the files are cached at the SBS as explained in Section \ref{sec:SBS}. In (b), the files are cached at the users as explained in Section \ref{sec:D2D}.
}
\label{fig:Cons_C}
\end{figure}

\begin{figure}
\centering
\begin{tikzpicture}
\begin{axis}[
scale=0.65,
grid,
ylabel={Energy consumption per Hertz($\mathrm{kJ}/Hz$)},
xlabel={ $\gamma$ },
title= (a) Caching at the SBS,
legend pos=outer north east
]

\addplot  table[x=C, y expr=\thisrow{noCache}/1000]{file3.dat};
\addplot  table[x=C, y expr=\thisrow{LFU}/1000]{file3.dat};
\addplot  table[x=C, y expr=\thisrow{LC}/1000]{file3.dat};
\addplot  table[x=C, y expr=\thisrow{PD}/1000]{file3.dat};
\addplot  table[x=C, y expr=\thisrow{Optimal}/1000]{file3.dat};

\addlegendentry{No caching};
\addlegendentry{LRU};
\addlegendentry{LCA};
\addlegendentry{PDCA};
\addlegendentry{Optimal solution};

\end{axis}
\end{tikzpicture}
\begin{tikzpicture}
\begin{axis}[
scale=0.65,
grid,
ylabel={Energy consumption per Hertz ($\mathrm{kJ}/Hz$)},
xlabel={ $\gamma$ },
title= (b) Caching at the user devices,
legend pos=outer north east
]

\addplot  table[x=C, y expr=\thisrow{noCache}/1000]{file4.dat};
\addplot  table[x=C, y expr=\thisrow{LFU}/1000]{file4.dat};
\addplot  table[x=C, y expr=\thisrow{LC}/1000]{file4.dat};
\addplot  table[x=C, y expr=\thisrow{PD}/1000]{file4.dat};
\addplot  table[x=C, y expr=\thisrow{Optimal}/1000]{file4.dat};

\addlegendentry{No caching};
\addlegendentry{LRU};
\addlegendentry{LCA};
\addlegendentry{PDCA};
\addlegendentry{Optimal solution};
\end{axis}
\end{tikzpicture}
\vspace{-1em}
\caption{ Energy consumption of the \ac{MBS} for different  values of the Zipf distribution parameter $\gamma$ ($\hat C = 10$, $U =3$). 
In (a) the files are cached at the SBS as explained in Section \ref{sec:SBS}. In (b) the files are cached at the users as explained in Section \ref{sec:D2D}.
}
\label{fig:Cons_gamma}
\end{figure}
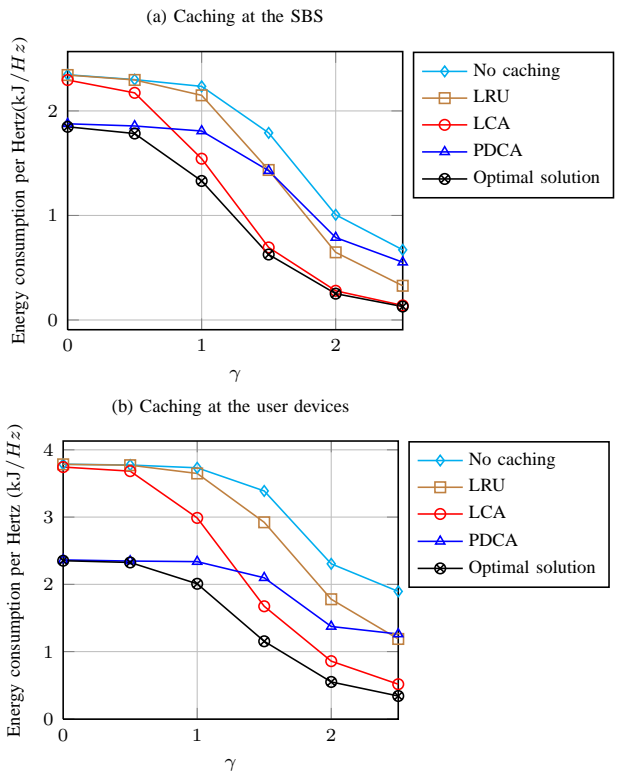

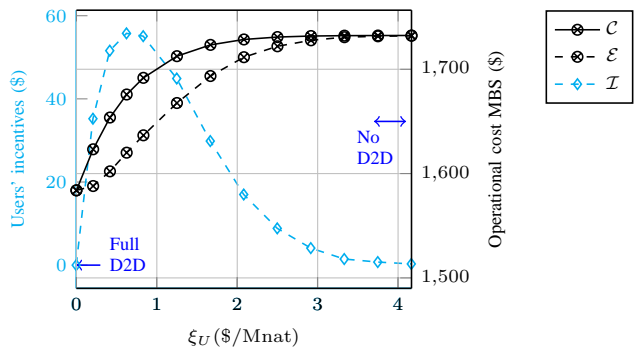
\begin{figure}
\centering
\begin{tikzpicture}
\begin{axis}[
scale=0.65,
color={cyan},
ylabel={Users' incentives ($\$$)},
axis y line*=left,
mark options={solid},
axis x line*=bottom
]
\addplot[cyan,mark=diamond, dashed]  table[x= XiUser, y = CostUSer]{EconomicalCost.dat}; 
\draw[font=\scriptsize, text width=0.5cm, blue,<-] (axis cs:0,0)-- (axis cs:0.3,0)node[right,yshift=0.15cm]{Full D2D};
\end{axis}
\begin{axis}[
mark options={solid},
scale=0.65,
grid,
axis y line*=right,
ylabel={Operational cost MBS (\$)},
xlabel={ $\xi_U (\mathrm{\$ / Mnat})$ },
color={black},
ymin = 1490,
legend style={at={(1.4,1)},anchor=north west}
]
\addplot[black,mark=otimes]  table[x= XiUser, y = TotalCost]{EconomicalCost.dat};
\addplot[black,mark=otimes, dashed]  table[x= XiUser, y = CostMBS]{EconomicalCost.dat}; 
\draw[font=\scriptsize, text width=0.5cm, blue,<->] (axis cs:3.7,1650)-- (axis cs:4.1,1650)node[below,xshift=-0.4cm]{No D2D};

\addlegendentry{$\mathcal{C}$}
\addlegendentry{$\mathcal{E}$}
\addlegendimage{cyan,mark=diamond, dashed}
\addlegendentry{$\mathcal{I}$}
\end{axis}
\end{tikzpicture}
\vspace{-1em}
\caption{Economical cost at the MBS for different economical incentives per transmitted data in the D2D links ($U=3$, $\gamma = 1$, $\hat C = 10$, and $\xi_{MBS}= 0.3 \$/KWh$).}
\label{fig:Cost}
\end{figure}

First, in Figs. \ref{fig:Cons_C}-\ref{fig:Cons_gamma}, we focus on the energy consumption at the MBS considering that the users
cooperate altruistically with the MBS (i.e., $\hat{g}_{uu'} = 0, \forall u, u\ne u$).
This allows us to fairly compare the SBS and D2D scenarios, and to evaluate the effect of having a centralized or distributed cache.
Afterwards, in Fig. \ref{fig:Cost}, we consider the cost of the MBS when users cooperate in exchange of an economical incentive.
Figs. \ref{fig:Cons_C}(a) and \ref{fig:Cons_C}(b) evaluate the energy consumption of the \ac{MBS} for different sizes of the cache capacity in the SBS and D2D scenarios, respectively.
It is observed that in both scenarios  the MBS energy consumption  decreases with the cache capacity for all the caching strategies.
When the   jointly  optimal transmission and caching strategy is compared with traditional non-caching solutions, it is observed that the optimal policy reduces the MBS energy consumption by 53.59\%  in the SBS scenario, and by 61.78\% in the D2D scenario.
This reduction is obtained 
when the cache capacity is 25\% of the average user traffic\ie $\hat C = 25$; however, further energy savings can be achieved by increasing the total cache capacity.
The performance of the online LRU caching policy is far from the optimal;
however, this was expected as it does not exploit any information about the future requests. 
Next, we assess the performance of policies that exploit only one of the caching gains, namely, pre-downloading (PDCA) or local caching gains (LCA).
Interestingly, it is observed  in the SBS scenario that  LCA achieves more energy savings than  PDCA (see Fig. \ref{fig:Cons_C}(a));
while in the D2D scenario,  PDCA  requires a lower energy consumption than  LCA (see Fig. \ref{fig:Cons_C}(b)).
This different behavior is later argued in the following paragraph as it depends on the value of $\gamma$.
Finally, if we globally compare the two scenarios, we observe that the deployment of SBSs leads to more reduction in the MBS energy consumption compared to the D2D scenario. 
The rationale behind this  is two-fold.
First, the expressions for  energy consumption in the objective functions of problems \eqref{eq:problemLinear} and \eqref{eq:problemLinearD2D} are different;
specifically, the energy consumption in the SBS scenario is always smaller (or equal) than the consumption in the D2D scenario, which can be proved by using Jensen's inequality.
Second, in the D2D scenario the total storage capacity is distributed across users instead of being centralized and, as a result, the energy consumption increases.

Figs. \ref{fig:Cons_gamma}(a) and \ref{fig:Cons_gamma}(b) evaluate the 
impact of the file popularity distribution, which is controlled by $\gamma$, on the energy consumption of the \ac{MBS} for the scenarios and algorithms mentioned above.
We observe that the energy consumption is dramatically reduced when  
$\gamma$ increases (and thus, 
the  popularity distribution becomes skewed) as more file repetitions are encountered.
By focusing on the policies \textit{No caching} and \textit{PDCA}, we observe that both follow a similar  trend when $\gamma$ increases;
the energy consumption of these policies decreases 
with $\gamma$ because it is more likely that two users request the same file within the same time slot.
Indeed, for the case of a single user connected to the SBS, $U = 1$, it can be observed that the energy consumption of these policies is not affected by $\gamma$.
Under a uniform popularity distribution of the files ($\gamma = 0$),  \ac{PDCA} outperforms  \ac{LRU} and \ac{LCA} as file repetitions are unlikely;
however, there is a crossing point between the performances of these policies. 
Interestingly, this crossing point occurs later in the D2D scenario, and it moves to larger values of $\gamma$ as the number of users increases.
We believe that this is because the more distributed the cache is, the more the local caching gain is penalized.
 In contrast, the pre-downloading gain is not affected much by distributing the cache across users.
Note that the optimal policy outperforms all the other policies by adapting to the best available gain for any value of $\gamma$.

In the previous simulations we have considered that users  altruistically cooperate  with the MBS. Next, we consider that users transmit over the D2D links
in exchange of an economical incentive of $\xi_{U}$ dollars per transmitted data. In this context, we minimize  
the economical cost at the MBS defined as the sum of the electricity bill, $\mathcal{E}$, and the incentives paid to users $\mathcal{I}$\ie  $\mathcal{C} = \mathcal{E} + \mathcal{I}$, where $\mathcal{E}= \int_{0}^{T} \xi_{MBS}\left (\sum_{u=1}^U W/U(\exp (\rateUser(\tau) U / W ) -1)\right ) \d \tau$ and 
$\mathcal{I} =  \int_{0}^{T} \xi_{U}\sum_{u=1}^U \sum_{u'\neq u}  \rateUserUser[\tau,u']{u}  \d \tau $.
We have set $\xi_{MBS}= 0.3 \mathrm{~\$/KWh}$, which is a typical electricity price.
Fig. \ref{fig:Cost} depicts the costs obtained under the optimal transmission and caching policy as 
$\xi_{U}$ varies in the $x$-axis.
The plot  $\mathcal{I}$ refers to the left $y$-axis, whereas the plots $\mathcal{C}$ and $\mathcal{E}$ refer to the right $y$-axis.
 The D2D links are used at its greatest possible extent when $\xi_{U}=0$ as D2D transmissions do not incur any cost to the MBS.
 Then, the users incentives first increases with $\xi_{U}$ until it starts decreasing as the usage of the D2D links is reduced. Finally, for very large 
values of $\xi_{U}$, the MBS serves all the traffic and no data is transmitted over the D2D links.

\vspace{-1em}
\section{Conclusions}
\label{sec_Conclusions}
This paper has investigated the opportunities that caching offers to reduce a generic cost function of the transmission rates\eg the required energy, bandwidth, or traffic to serve the users.
Two different scenarios have been considered where the information is either cached at an SBS, or directly at the user terminals, which then use D2D communications to share the cached contents.
It has been shown that, when the transmission and caching policies are jointly designed,
 the cache offers two possible gains, namely, the \textit{pre-downloading} and \textit{local} caching gains.
In both scenarios, 
the jointly optimal transmission and caching policy has been obtained by 
 demonstrating that constant rate caching within a time slot is optimal, which allows to reformulate the infinite-dimensional optimization problem as a solvable convex program.
The numerical results have focused on  minimizing the energy consumption at the MBS.
It has been shown that the proposed solutions achieve
 substantial energy savings.
Specifically, when the cache capacity is only 25\% of the average traffic of a single user, energy savings of more than 53\% have been obtained.
It has been observed that the pre-downloading gain is greater than the local caching gain when the file popularity distribution is uniform, and vice versa when the file popularity is skewed.
The proposed optimal offline transmission and caching policies can be used as a lower bound to evaluate the cost of  any online policy.
In particular, it has been observed that, in the considered wireless setting, the popular LRU online algorithm performs far from the optimal strategy as it does not exploit the pre-downloading caching gain.
To conclude, our results motivate the design of novel online algorithms that can better approach the performance of the optimal offline solution, which is   left for future work.
These algorithms can be designed by exploiting the partial knowledge of some of the subsequent file requests (e.g., when some users are watching long video content that span several time slots), or by learning  users' daily  behaviors.

\setcounter {section} {0}
\def\thesubsection{\Alph{subsection}}
\def\thesubsubsection{\Alph{subsection}-\arabic{subsubsection}}

\def\thesubsection{\Alph{subsection}}
\def\thesubsubsection{\Alph{subsection}-\arabic{subsubsection}}

\vspace{-1em}
\section*{Appendix}

In the  statement of Lemma \ref{lemma:constantCachingSBS}, we assume that we know the optimal number of data units to be cached for each request, $\cached^\star$.
Thus, we also know the optimal value of the maximum and minimum data departure curves at slot transitions\ie
$b_{n} \triangleq B(n\Ts,\cachingP^\star) =  C +\sum_{\ell = 1}^{n} \sum_{u=1}^{U} \Ts \servingNU[n u] -  \cached^\star $ and
$a_{n} \triangleq A(n\Ts,\cachingP^\star)= \sum_{\ell = 1}^{n} \sum_{u=1}^{U}  \Ts\servingNU[n u] - \cached[\bar{\rhoFunction}(n,u)]^\star$ (c.f. Definitions \ref{def:maxD} and \ref{def:minD}).
$\bar{\rhoFunction}(n,u)$ is defined after the problem in \eqref{eq:problemLinear}.
However, we do not know the actual values of $B(t,\cachingP^\star)$ and $A(t,\cachingP^\star)$ for $t \neq n \Ts$ since it depends on the  shape of the optimal caching policy.


We first relax the problem  in  \eqref{eq:problem}
by considering  the constraints  in \eqref{eq:c2} and \eqref{eq:c1} only at slot transitions ($t = n\Ts$, $\forall n$).
Thus, we consider the following relaxed problem:
\begin{subequations}
\label{eq:problemAppendix}
\begin{align}
\minimize_{ \{ r(t), \caching \}_{t\in[0,T]}} &  \int_{0}^{T} g (r(\tau)) \d \tau &\\
 \st\quad \quad & a_n \leq \departure[nTs] \leq b_n,   &\hspace{-2.7cm} \forall n, \\
  & r(t) \geq 0,  &\hspace{-2.7cm} \forall t \in [0,T],  \\
  &\vec 0 \preceq \caching \preceq \serving, &\hspace{-2.7cm}  \forall t \in [0,T], \label{eq:Apc4} \\
  &B(n\Ts,\cachingP) = b_n,  A(n\Ts,\cachingP) = a_n, &    \forall n, \label{eq:Apc5}
\end{align}
\end{subequations}
where the values of  $a_n$ and $b_n$ are known $\forall n$  as argued above.

Note that any caching policy, $\cachingP$, satisfying \eqref{eq:Apc4}-\eqref{eq:Apc5} is optimal to the relaxed problem.

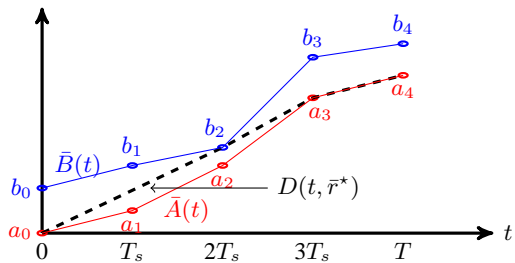
\begin{figure}
\centering
\begin{tikzpicture}[
	yscale = 0.6,
	xscale = 1.2,
    axis/.style={very thick, ->, >=stealth'},
    important line/.style={thick},
    dashed line/.style={dashed, thin},
    pile/.style={thick, ->, >=stealth', shorten <=2pt, shorten
    >=2pt},
    every node/.style={color=black,font=\small}
    ]
    \draw[axis] (0,0)  -- (5,0) node(xline)[right] {$t$};
    \draw[axis] (0,0)  -- (0,5) node(xline)[left] {};

    \node[below] at (0,0){0};
    \node[below] at (1,0){\Ts};
    \node[below] at (2,0){2\Ts};
    \node[below] at (3,0){3\Ts};
	\node[below] at (4,0) {$T$}; 

	\draw[red,thick, solid] (0,0) circle (0.05cm) node[left, red]{$a_0$};
	\draw[red,thick, solid] (1,0.5) circle (0.05cm) node[below, red]{$a_1$};
	\draw[red,thick, solid] (2,1.5) circle (0.05cm) node[below, red]{$a_2$};
	\draw[red,thick, solid] (3,3) circle (0.05cm) node[below, red, xshift=0.1cm]{$a_3$};
	\draw[red,thick, solid] (4,3.5) circle (0.05cm) node[below, red]{$a_4$};

	\draw[blue,thick, solid] (0,1) circle (0.05cm) node[left, blue]{$b_0$};
	\draw[blue,thick, solid] (1,1.5) circle (0.05cm) node[above, blue]{$b_1$};
	\draw[blue,thick, solid] (2,1.9) circle (0.05cm) node[above, blue, xshift=-0.1cm]{$b_2$};
	\draw[blue,thick, solid] (3,3.9) circle (0.05cm) node[above, blue]{$b_3$};
	\draw[blue,thick, solid] (4,4.2) circle (0.05cm) node[above, blue]{$b_4$};	
	
	\draw[dashed, very thick] (0,0)-- (2,1.9) --(3,3)-- (4,3.5);
	\draw[red, thin] (0,0)-- (1,0.5) --(2,1.5)-- (3,3)--(4,3.5);
 
	 \draw[blue, thin] (0,1)-- (1,1.5) --(2,1.9)-- (3,3.9)--(4,4.2);

 		\node[blue, align=center] at (0.4,1.5){$\bar B(t)$};
		\node[red] at (1.6,0.5){$\bar A(t)$};
		
		\draw[->] (2.5,1) node[right]{$D(t, \bar r^\star)$} --(1.2,1);

\end{tikzpicture}
\vspace{-1em}\caption{Representation of the relaxed problem in \eqref{eq:problemAppendix}.}
\label{fig:proofSjaping}
\end{figure}

This problem is represented in Fig. \ref{fig:proofSjaping}.
Define  $\bar r^\star(t)$
as the optimal transmission rate
to the relaxed problem in \eqref{eq:problemAppendix}.
The optimal data departure curve, $D(t, \bar r^\star)$, is a piece-wise linear function that can be
obtained as the tightest string whose ends are tied to $(0,0)$ and $(0, a_{N})$.
This statement is proved in  \cite{zafer_calculus_2009} by using the integral version of the Jensen's inequality and  the convexity of the cost function $g(\cdot)$.
Accordingly, for any feasible caching policy $\vec c$, we know that the transmission rate of this relaxed problem might only change at slot transitions.
Due to this, the optimal data departure curve to the problem in  \eqref{eq:problemAppendix}, $ D(t, \bar r^\star)$, 
satisfies
$\bar A(t) \leq   D(t, \bar r^\star)\leq \bar B(t)$, $\forall t \in [0, T]$, where
$\bar A(t)$ is the piece-wise linear curve obtained by joining the points $(n\Ts,  a_{n})$ for $n=0,\dots,N$; and
$\bar B(t)$ is the piece-wise linear curve obtained by joining the points $(n\Ts,  b_{n} )$ for $n=0,\dots,N$ (see Fig. \ref{fig:proofSjaping}).

Next consider the caching policy that caches each file at a constant rate, \optimalCaching[], as defined in Lemma \ref{lemma:constantCachingSBS}.
First note that \optimalCaching[] satisfies \eqref{eq:Apc4}-\eqref{eq:Apc5}; and thus, it is an optimal caching policy of the relaxed problem in \eqref{eq:problemAppendix}.
Additionally, the maximum and minimum data departure curves associated to the caching policy  \optimalCaching[] are piece-wise linear and satisfy
\begin{equation}
\label{relaxed}
A(t,\optimalCaching[])= \bar A(t) \leq D(t, \bar r^\star)\leq \bar B(t) = B(t,\optimalCaching[]), \forall t\in [0, T].
\end{equation}
Thus, $\{\bar r^\star, \optimalCaching[]\}$ is the optimal solution to the relaxed problem in \eqref{eq:problemAppendix}.

Note that, from \eqref{relaxed}, the pair $\{\bar r^\star, \optimalCaching[]\}$ satisfies the constraints that had been relaxed in the original problem in \eqref{eq:problem}. 
Accordingly, since  $\{\bar r^\star, \optimalCaching[]\}$ is a feasible solution to the original problem, 
 and the objective function is the same in both problems,
it is also an optimal solution.
\hfill\IEEEQED

\vspace{-1em}
\bibliographystyle{IEEEtran}
\bibliography{IEEEabrv,Bib_Maria_centralized}

\end{document}